\documentclass[twocolumn, amsmath,amssymb,aps,pra,superscriptaddress]{revtex4-2}

\usepackage{graphicx, xcolor}
\usepackage{braket}
\usepackage{subcaption}

\newcommand{\figref}[1]{\figurename~\ref{#1}}
\newcommand{\secref}[1]{Sec.~\ref{#1}}

\renewcommand{\eqref}[1]{Eq.~(\ref{#1})}

\newcommand{\ip}{\mathrm{I}_\mathrm{p}}
\newcommand{\up}{\mathrm{U}_\mathrm{p}}
\newcommand{\pb}{\mathbf{p}}

\newcommand{\rb}{\mathbf{r}}

\newcommand{\Ab}{\mathbf{A}}

\renewcommand\Re{\mathrm{Re}}
\renewcommand\Im{\mathrm{Im}}
\newcommand{\sign}{\mathrm{sign}}

\begin{document}

\title{Dissecting Sub-Cycle Interference in Photoelectron Holography}
\author{Nicholas Werby}
\thanks{These two authors contributed equally}
\affiliation{Stanford PULSE Institute, SLAC National Accelerator Laboratory\\
2575 Sand Hill Road, Menlo Park, CA 94025, USA}
\affiliation{Department of Physics, Stanford University, Stanford, CA 94305, USA}

\author{Andrew S. Maxwell}
\thanks{These two authors contributed equally}
\affiliation{Institut de Ciencies Fotoniques, The Barcelona Institute of Science and Technology, 08860 Castelldefels (Barcelona), Spain}
\affiliation{Department of Physics \& Astronomy, University College London, Gower Street,  London,  WC1E 6BT, United Kingdom}

\author{Ruaridh Forbes}
\affiliation{Stanford PULSE Institute, SLAC National Accelerator Laboratory\\
2575 Sand Hill Road, Menlo Park, CA 94025, USA}
\affiliation{Department of Physics, Stanford University, Stanford, CA 94305, USA}
\affiliation{Linac Coherent Light Source, SLAC National Accelerator Laboratory, Menlo Park, California 94025, USA}
\author{Philip H. Bucksbaum}
\affiliation{Stanford PULSE Institute, SLAC National Accelerator Laboratory\\
2575 Sand Hill Road, Menlo Park, CA 94025, USA}
\affiliation{Department of Physics, Stanford University, Stanford, CA 94305, USA}
\affiliation{Department of Applied Physics, Stanford University, Stanford, CA 94305, USA}

\author{Carla Figueira de Morisson Faria}
\email{c.faria@ucl.ac.uk}
\affiliation{Department of Physics \& Astronomy, University College London, Gower Street,  London,  WC1E 6BT, United Kingdom}


\date{\today}

\begin{abstract}
Multipath holographic interference in strong-field quantum tunnel ionization is key to revealing sub-Angstrom attosecond dynamics for molecular movies. This critical sub-cycle motion is often obscured by longer time-scale effects such as ring-shaped patterns that appear in above-threshold ionization (ATI).  
In the present work, we overcome this problem by combining two novel techniques in theory and experimental analysis: unit-cell averaging and time-filtering data and simulations. 
Together these suppress ATI rings and enable an unprecedented highly-detailed quantitative match between strong-field ionization experiments in argon and the Coulomb-quantum orbit strong-field approximation (CQSFA) theory.
Velocity map images reveal fine modulations on the holographic spider-like interference fringes that form near the polarization axis. CQSFA theory traces this to
the interference of three types of electron pathways. 
The level of agreement between experiment and theory allows sensitive determination of quantum phase differences and symmetries, providing an important tool for quantitative dynamical imaging in quantum systems. 
\end{abstract}


\maketitle
\section{Introduction}
\label{sec:Intro}

In the study of attosecond ($10^{-18}$~s) science, probing matter with a strong laser field has emerged as a prominent tool for revealing internal dynamics of atoms and molecules \cite{faria_it_2020,huismans_time-resolved_2011,Zuo1996,Itatani2004}. The photoelectron emitted in strong-field ionization (SFI) can follow a wide variety of field-driven trajectories depending on the phase of the laser field at the time of ionization. 
Photoelectron vector momentum distributions (PMD) encode these trajectories as intricate interference patterns displayed in angularly resolved photoelectron measurements. Significant work has been applied towards isolating and disentangling these patterns in order to determine the electron \cite{huismans_time-resolved_2011,Bian2011, Bian2012,hickstein_direct_2012,haertelt_probing_2016,Lai2017, Maxwell2017, Maxwell2017a, Maxwell2018, he_direct_2018, Kubel2019, Kang2020, Maxwell2020, werby_disentangling_2021} and sometimes the core \cite{Veltheim2013,Mi2017,walt_dynamics_2017} dynamics. 

\begin{figure}
    \centering
    \begin{subfigure}{0.99\linewidth}
    \includegraphics[width=\linewidth]{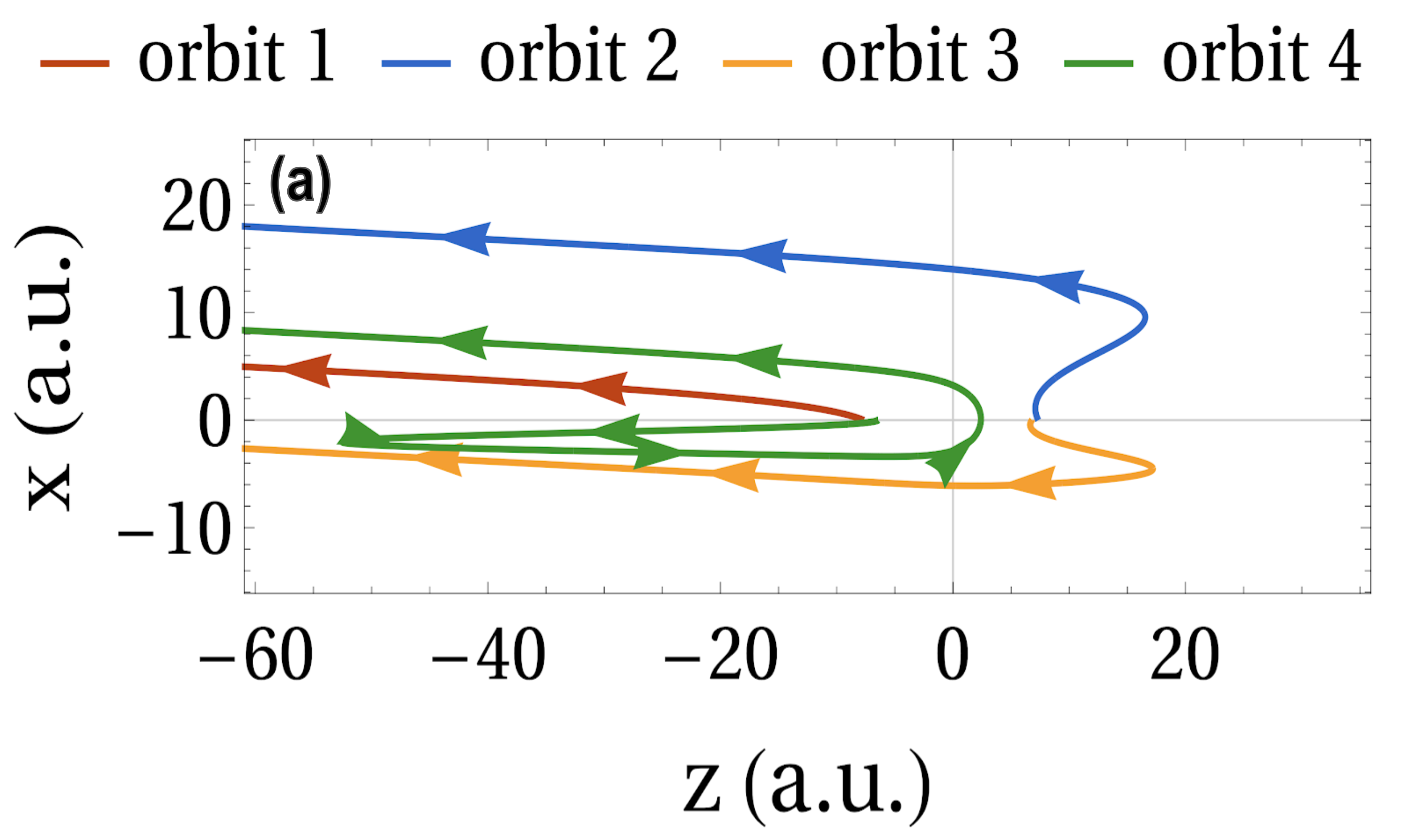}
    \end{subfigure}
    \begin{subfigure}{0.99\linewidth}
    \includegraphics[width=\linewidth]{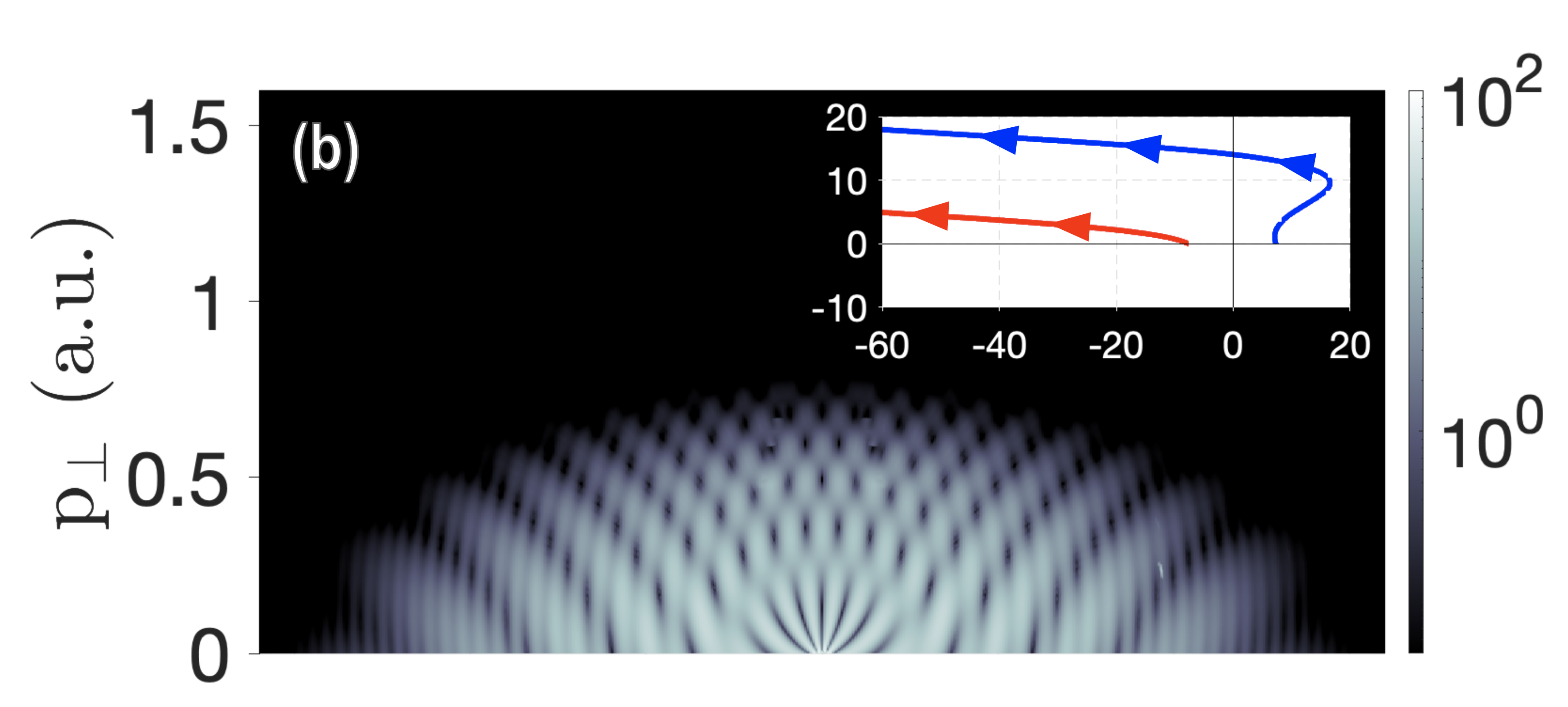}
    \end{subfigure}
    \begin{subfigure}{0.99\linewidth}
    \includegraphics[width=\linewidth]{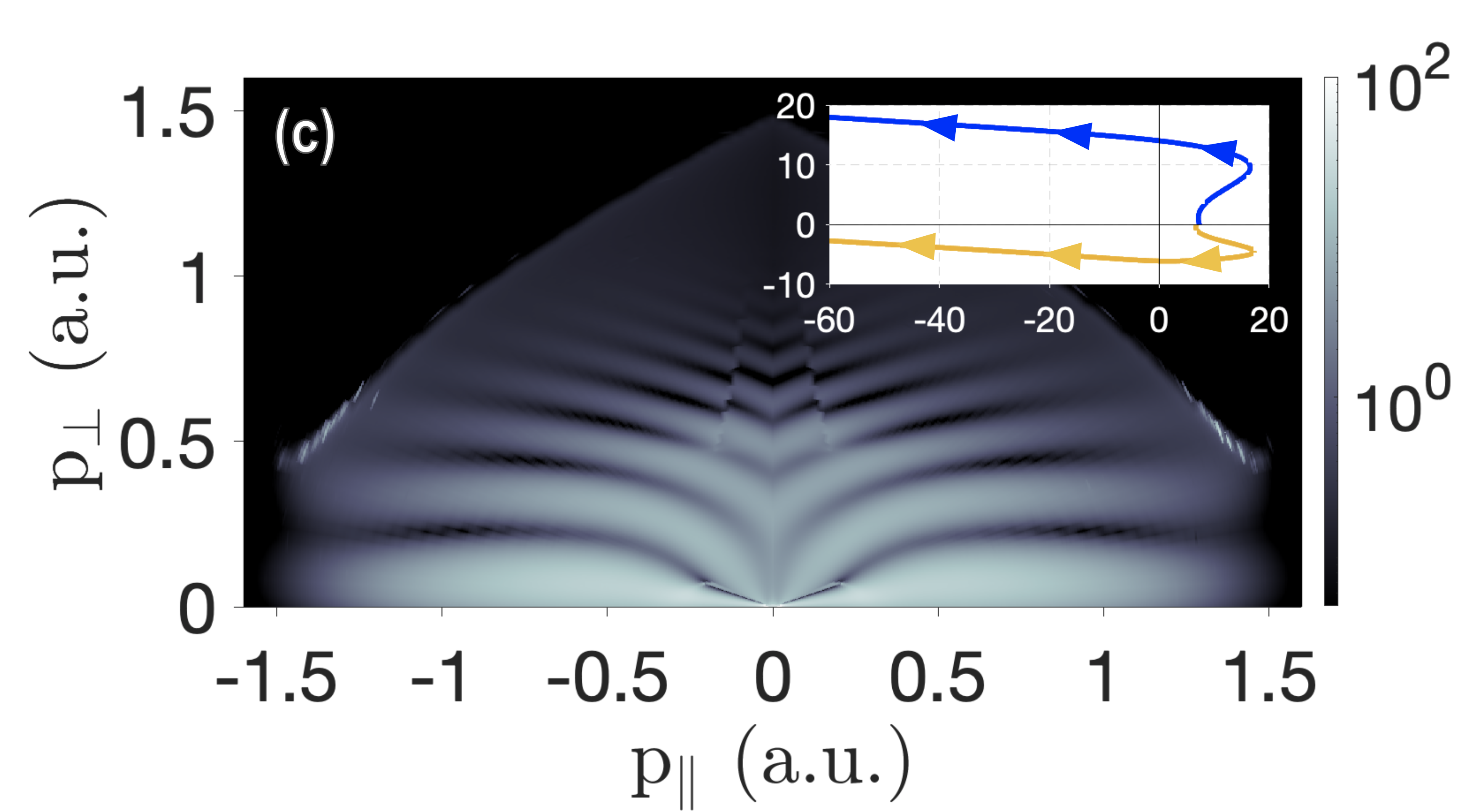}
    \end{subfigure}
    \caption{(a) The four CQSFA trajectories found by solving the saddle point equations \eqref{eq:s1} and \eqref{eq:s2}. The arrows mark the direction of travel and passage of time: between each arrow head $0.2$ cycles pass. The trajectories correspond to the final momentum $\pb=(-1.0, 0.13)$. (b) The interference between orbits 1 and 2 produces the fan structure, and (c) the interference between orbits 2 and 3 produce the spider-leg structure. Here, both (b) and (c) have been unit-cell averaged, see Sec. \ref{sec:unit_cell_averaging}.}
    \label{fig:CQSFATrajecotories}
\end{figure}

The interference of photoelectron trajectories contains information about the structure of the underlying parent ion, and a breakthrough in disentangling PMDs to probe the parent atom or molecule came in the form of photoelectron holography \cite{huismans_time-resolved_2011, hickstein_direct_2012,faria_it_2020}. Ultrafast photoelectron holography brings together high electron currrents, coherence, and subfemtosecond resolution, and allows the retrieval of quantum phase differences. This makes it a popular alternative to pump-probe interferometric schemes such as the Reconstruction of Attosecond Burst By Interference of Two-photon Transition (RABBITT) technique \cite{Haessler2010}, the Spectral
Phase Interferometry for the Direct Electric Field Reconstruction (SPIDER) \cite{Cormier2005} and the Frequency Resolved Optical Gating (FROG) \cite{Gagnon2008} (for a review see Ref.~\cite{Orfanos2019}). The patterns visible in experiment are produced by the interference of different electronic pathways to the detector (see FIG.~\ref{fig:CQSFATrajecotories}~(a)). These pathways undergo varying degrees of interaction with the parent ion and so they pick up different phases. The interference between the trajectories, recorded by the detector, can reveal these phases and be employed for imaging. Many interference patterns have been identified as the combination of two photoelectron pathways which have been used to probe and image the target. Among these two-trajectory interference patterns are the fan-like structure \cite{Rudenko2004b, Maharjan2006, Gopal2009, Arbo2006, Lai2017} (see \figref{fig:CQSFATrajecotories}~(b)), the result of the interference between direct and forward deflected trajectories;
the spider-leg structure \cite{huismans_time-resolved_2011,Bian2011, hickstein_direct_2012} (see \figref{fig:CQSFATrajecotories}~(c)), the result of the interference between forward scattered and forward deflected trajectories;
 and the fishbone-like structure \cite{Bian2012, haertelt_probing_2016}, which occurs in the same region as the spider but has fringes that are nearly orthogonal to the polarization axis. 
All the holographic structure and analysis to date has relied on two-trajectory interference. However, many of the above-stated patterns overlap, with some models predicting at least four relevant trajectories (see \figref{fig:CQSFATrajecotories}~(a)) \cite{Maxwell2018}, while patterns like the fish-bone structure require elaborate experimental methodologies \cite{haertelt_probing_2016} to extract and differentiate from more dominating features. More preferable would be to use a multi-trajectory analysis \cite{Maxwell2017}. In this work we do just that presenting a three-trajectory pattern that leads to a `modulated spider'. 


A prominent technique to disentangle different types of quantum interference is to simulate combinations of electron trajectories and compare the calculation to an experimentally measured PMD. In principle, selectively turning on and off different trajectories within these simulations should yield holographic structures which can be matched to the experimental patterns. This analysis technique has been somewhat successful for some well-known holographic structures such as the `spider legs' described above \cite{huismans_time-resolved_2011, hickstein_direct_2012};
however, for holographic structures produced through the interference of electron trajectories which are significantly affected by the Coulomb potential of the parent ion there has been a mismatch between experiment and theory \cite{Rudenko2004b,Maharjan2006,quan_classical_2009,blaga_strong-field_2009,huismans_time-resolved_2011,Lai2017,faria_it_2020}. Many models of photoelectron holography address the potential via a Born series, which fails to converge well for long-range potentials  \cite{Spanner2004,huismans_time-resolved_2011,Bian2011,Bian2012}. The development of Coulomb-distorted quantum orbit models \cite{faria_it_2020} permits the inclusion of many previously neglected Coulomb effects. The Coulomb-quantum orbit strong-field approximation (CQSFA) \cite{Lai2015,Lai2017,Maxwell2017,Maxwell2017a,Maxwell2018,Maxwell2018PRA} is one such model. It provides a very clear picture with four interfering `orbits', which may be switched on and off at will. This has enabled computationally fast and accurate analyses of experimental features \cite{Kang2020, Maxwell2020}.

\begin{figure*}
    \centering
    \begin{subfigure}[b]{0.32\linewidth}
        \includegraphics[width=\linewidth]{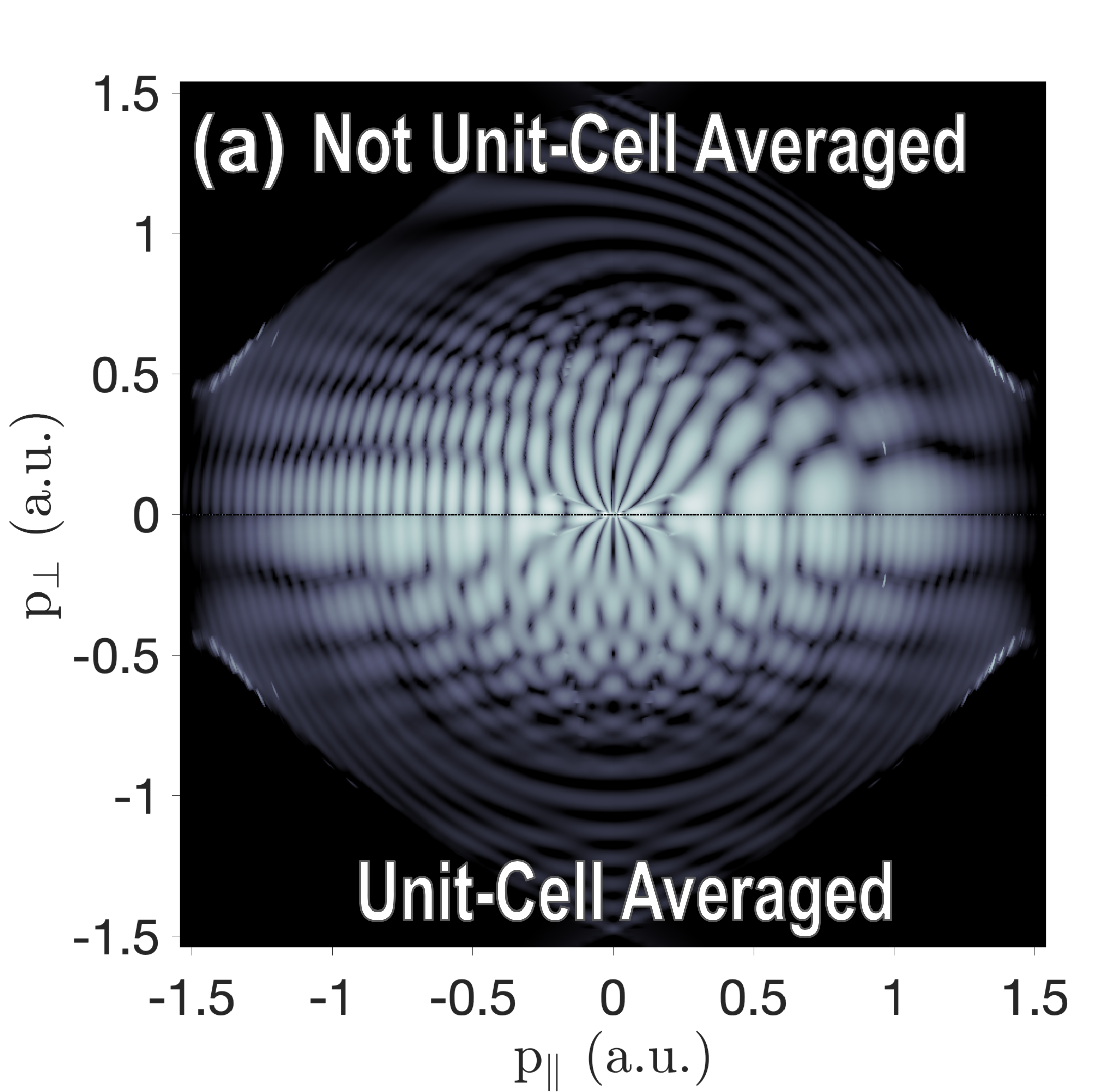}
    \end{subfigure}
    \begin{subfigure}[b]{0.32\linewidth}
        \includegraphics[width=\linewidth]{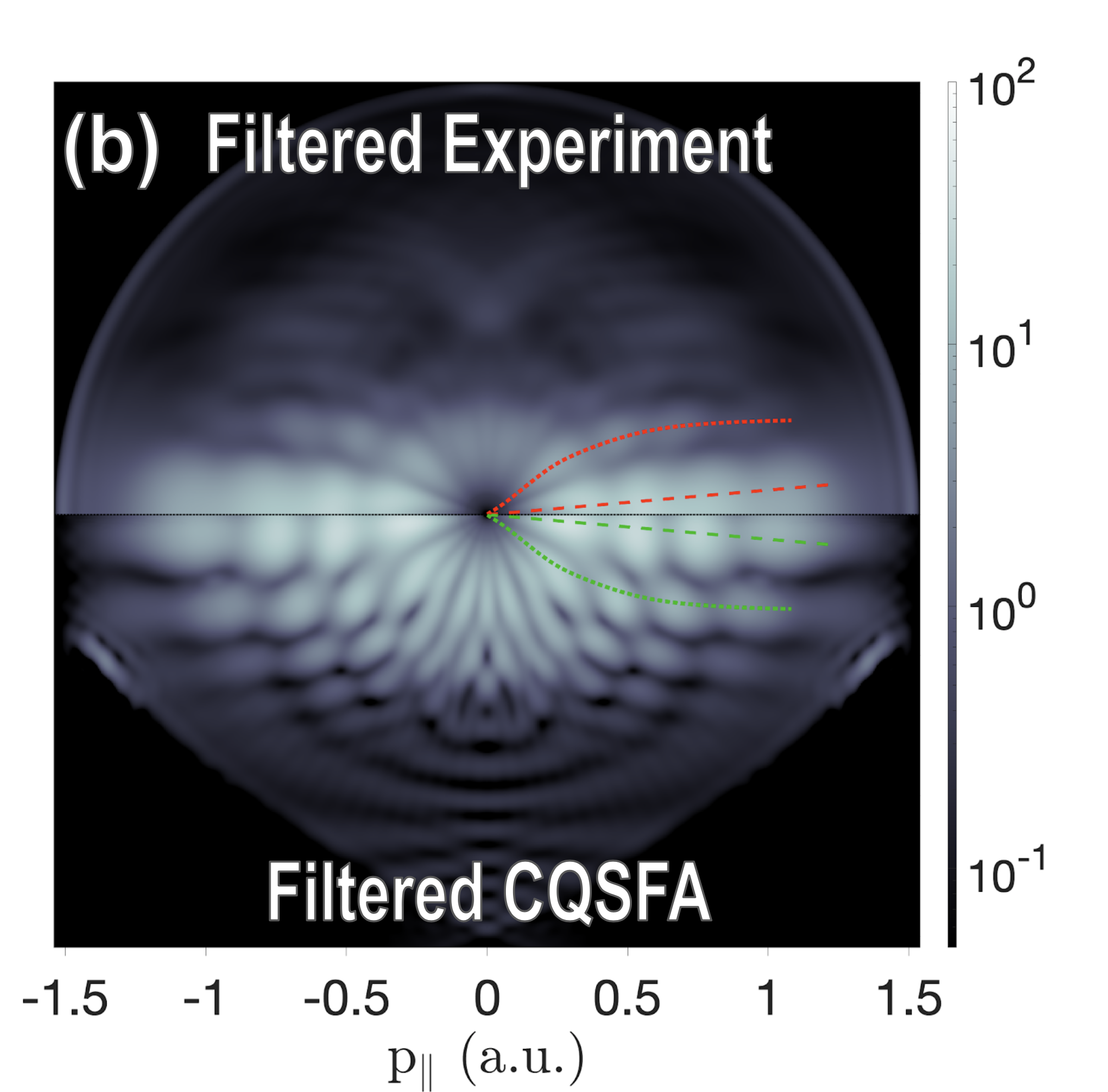}
    \end{subfigure}
    \begin{subfigure}[b]{0.32\linewidth}
        \includegraphics[width=\linewidth]{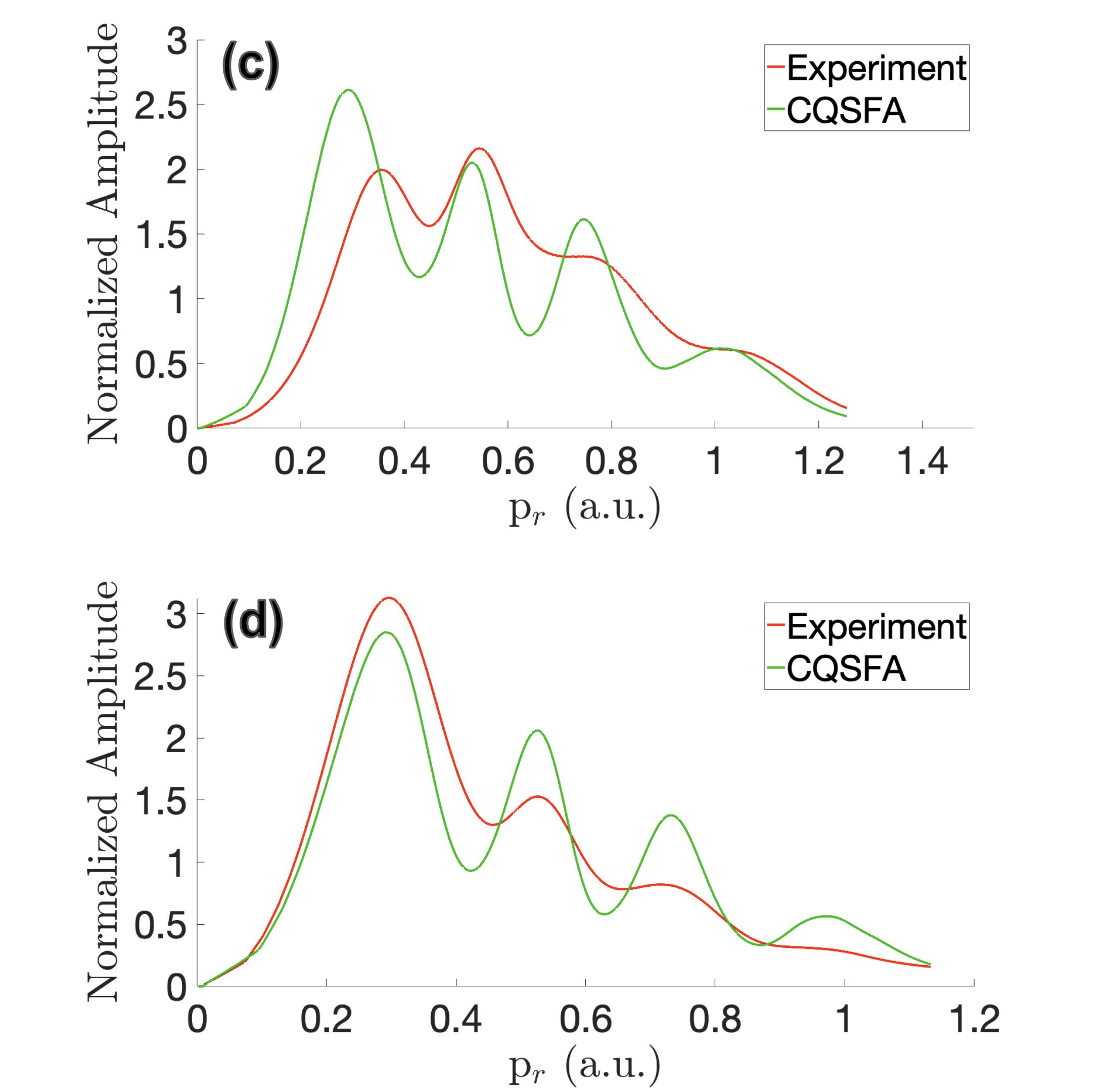}
    \end{subfigure}
    \caption{Photoelectron momentum dependent yield for argon for a laser intensity of $2\times 10^{14}$~W/cm$^2$ and wavelength of $\lambda=800$~nm. 
    Panel (a), bottom half shows the CQSFA with unit-cell averaging, while the top half shows without (the unit start is defined by taking $\phi=0$ in \eqref{eq:AField}).
    The top half of panel (b) shows the time-filtered experimental results, while the bottom half presents the unit-cell averaged CQSFA calculation after receiving the same filtering treatment as the experimental data.  Panels (c) and (d) respectively show the lineouts along close to the parallel axis and along the first spider leg as indicated on panel (b).}
    \label{fig:MainFig}
\end{figure*}

However, detailed multi-trajectory analyses of sub-cycle structures is only possible by overcoming the gap that forms between CQSFA calculations and experimental PMDs caused by certain theoretical and experimental barriers. Any model that imposes restrictions on the ionization times faces a technical problem for predicting sub-cycle phenomena, which causes it to diverge from experimental PMDs. Calculations can easily restrict ionization to a single-laser-cycle unit cell, thus focusing only on sub-cycle interference; however, this leads to simulated spectra with artificial asymmetries governed by the arbitrarily chosen start and end phases of the specific unit cell employed (see Sec.~\ref{sec:unit_cell_averaging}). Simply incorporating a longer unit cell (by including ionization events from more laser cycles) into the calculation to eliminate the asymmetry results in ringlike fringes from compounding above-threshold ionization (ATI) which obscure sub-cycle features and impede analysis. We encounter a similar barrier in the laboratory. In experiments in which the ionizing laser pulses contain approximately ten or more optical cycles, the resultant PMDs are dominated by significant ATI rings, which further disrupts analysis. Experimental attempts to eliminate ATI structures by moving to much shorter pulses leads to several other problems for analysis with the CQSFA. One- or two-cycle laser pulses are now possible in the laboratory. Unfortunately, these ultrashort pulses do not possess a single uniform electric field cycle, but rather have a significant time-varying field envelope, which generates a carrier envelope phase (CEP) parameter that governs the electron dynamics \cite{kling_imaging_2008, Bergues2012}. 
This parameter is not included in uniform cycle calculations, preventing straightforward comparisons with theory.
Furthermore, ultrashort pulses of this kind will suppress any holographic features that take more than a single cycle to form and unequal cycles may blur the patterns \cite{Maxwell2018, ShaoGang2020}.


In order to effectively investigate sub-cycle structures, we employ two novel techniques in the experimental analysis and the theoretical computation that bridge this gap between them. In the experimental analysis, a time-filtering technique is applied which effectively extracts sub-cycle information from spectra generated from multi-cycle laser pulses by eliminating the energy-periodic background generated by the ATI rings \cite{werby_disentangling_2021}. In the CQSFA calculation, unit-cell averaged computations are performed, in which the start and end points of the unit cell of ionizations are averaged over. This not only removes the aforementioned asymmetries but also ensures all combinations of trajectories that were present in the experiment are accounted for.

By bridging the gap between experiment and calculation, many previously unexplored subtle sub-cycle features are revealed. In this paper, we present a high fidelity PMD of argon gas photoionized by a multi-cycle laser pulse and filtered to remove the ATI dependence. We introduce the idea of unit-cell averaging in CQSFA calculations and demonstrate how it matches the experiment. 
Unit-cell averaging employs an ansatz which incoherently averages over ensembles of trajectories with different time ordering. Variations in the time ordering results from different initial conditions of the laser field, which accurately approximates the incoherent averaging that will occur in an experiment.
We then compare our experimental PMD to unit-cell averaged and filtered CQSFA calculations, and explore the newly revealed holographic features which are well matched between calculations and experiment. 

This article is organized in the following way. In \secref{sec:MeetingInTheMiddle} we compare experiment and theory with the methods of time-filtering and unit-cell averaging which enable effective comparison between sub-cycle features. Next, in \secref{sec:Filtering} and \ref{sec:unit_cell_averaging} the methodology of time-filtering and unit-cell averaging, respectively, is outlined. Following this, in \secref{sec:orbits} we demonstrate with the CQSFA that the modulations on the spider legs are a three-trajectory interference pattern. In \secref{sec:phases} this interference pattern is used to demonstrate the existence of Gouy and bound-state phases for the photoelectrons. Finally, in \secref{sec:conclusions} we state our conclusions.

\section{Sub-Cycle Interference Comparison: Bridging the Gap}
\label{sec:MeetingInTheMiddle}
The result of our experimental and theoretical efforts, with specific emphasis on the sub-cycle interference is shown in \figref{fig:MainFig}. 
In \figref{fig:MainFig}~(b) we show a high resolution, time-filtered experimental PMD of argon and compare with computations using the CQSFA. In general we find very strong agreement between the experiment and the CQSFA. The main features of the spider and fan-like structures are all clearly visible. Particularly good agreement is found near the polarization axis for the axial fringes and first spider leg.
Notable features in the experimental spectra are modulations on the spider legs (see dotted and dashed lines), which are visible due to the exceptionally high resolution of the experiment, while the time-filtering technique separates and highlights the modulations with fringes that are broader than the ATI rings. These modulations are well-matched by the unit-cell averaged and filtered CQSFA calculation, \figref{fig:MainFig}~(b).

\begin{figure*}
    \centering
    \begin{subfigure}[b]{0.49\linewidth}
        \includegraphics[width=\linewidth]{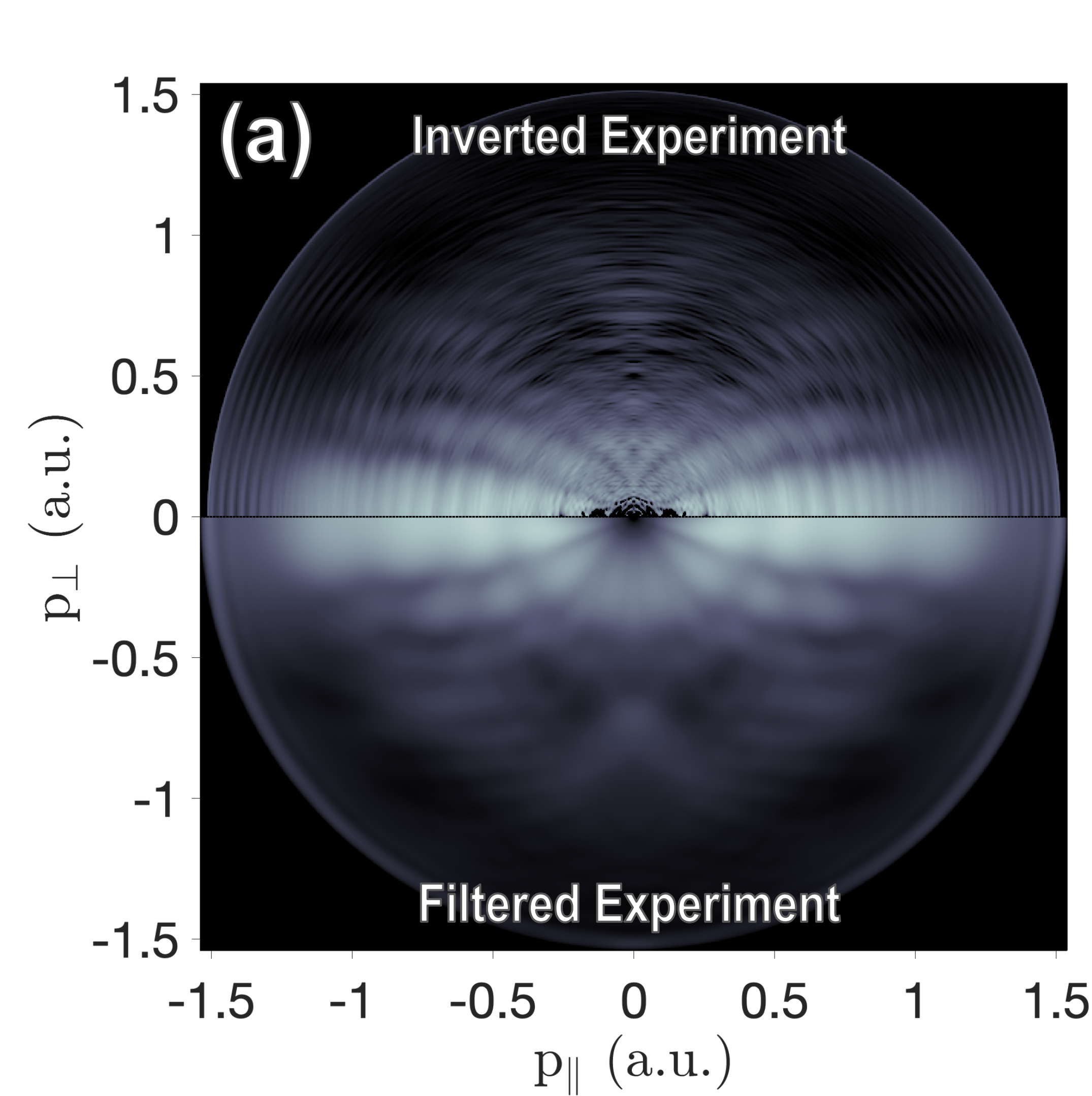}
    \end{subfigure}
    \begin{subfigure}[b]{0.49\linewidth}
        \includegraphics[width=\linewidth]{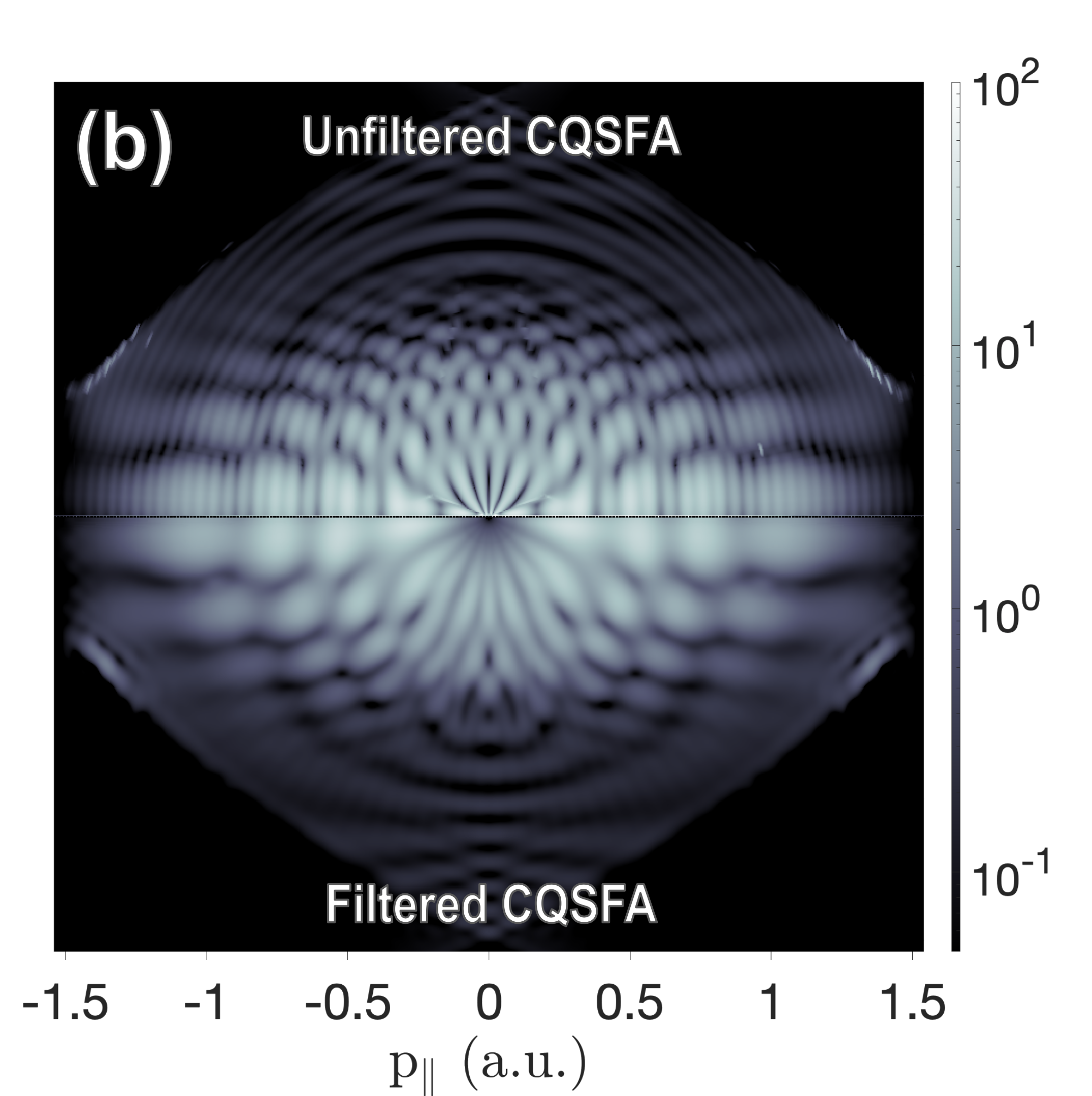}
    \end{subfigure}
    \caption{The application of the time-filtering technique to both the experimental data and the CQSFA calculations. (a) The top half shows the raw experimental data after it has been inverted via polar onion-peeling. (b) The top half shows the CQSFA calculation with just unit-cell averaging applied (see \secref{sec:unit_cell_averaging}). The bottom halves of both panels display the result of the time-filtering upon the top halves. The ATI rings and the fine modulations of the CQSFA have both been removed, without disrupting underlying structure.}
    \label{fig:Filtering}
\end{figure*}

In panel (a) of \figref{fig:MainFig} we present the CQSFA results with and without unit-cell averaging and without any filtering. Without the unit-cell averaging the CQSFA results are asymmetric (see \secref{sec:unit_cell_averaging} and Appendix \ref{appendix:unit_cell_averaging} for more details) and the modulations along the spider legs are not correctly reproduced. However, on the lower right-hand [left-hand] side of the panel broad [fine] modulations on the spider legs can be seen. It is a combination of both the broad and fine modulations (unit-cell averaging incoherently mixes both sides of the PMD) that leads to the modulation seen in experiment. 
Fine modulations are visible in the inverted experimental data;
(see \figref{fig:Filtering} and \secref{sec:Filtering} for more details)
however, it is not clear whether these interferences trace back to these CQSFA fine modulations or to the ATI rings. Filtering both the experiment and the CQSFA data removes the fine modulations and the ATI rings and thus allows for an unambiguous comparison of the two.

In panels (c) and (d) we plot lineouts near the parallel axis (i.e. along the laser polarization direction) and along the first spider leg, respectively, from both the filtered experimental and filtered CQSFA results in panel (b) (see dotted and dashed lines). 
In both panels broad modulations along the axial and first spider leg lineout are observed. Both the period of modulation as well as the overall signal amplitude are in good agreement between the experimental and theoretical results, except at higher momenta. Only the modulation depth is not so well matched, which could be explained by incoherent effects such as variation of the laser intensity over the focal volume.

To fully analyze these results we must understand some further details on the time-filtering and unit-cell averaging techniques as well as certain details about the experimental and theoretical methods. 

\section{Experimental Methods and Time-Filtering}
\label{sec:Filtering}

We employ common techniques for the strong-field ionization of argon atoms. Argon gas is pulsed through an Even-Lavie \cite{even_even-lavie_2015} valve before being strong-field ionized by an 800~nm, 40~fs, linearly polarized Ti:sapphire laser pulse with 200~TW/cm$^2$ peak intensity. 
The intensity was determined by fitting the signal drop-off predicted by the CQSFA along the axial lineout (\figref{fig:MainFig}~(c)) to the experiment. 
Fits were performed at 25 TW/cm$^2$ intervals. In this way we conclude that our intensity is determined to no better than about 7\%. This value is consistent with a calculation based on measured focal parameters for the setup.

The photoelectrons are extracted in a velocity map imaging (VMI) spectrometer \cite{eppink_velocity_1997}, impact a micro-channel plate detector and phosphor screen, and are recorded by a CCD camera. On-the-fly peak finding \cite{chang_improved_1998} is employed to increase the fidelity of the final spectrum. For the experimental results shown here, 63 billion electron impacts are recorded.

The laser pulse is linearly polarized in the detector plane so the VMI records an axial and perpendicular projection of the cylindrically symmetric vector momentum for each electron. This may be inverted to generate the $p_\parallel - p_\perp$ cross section of the ionized Newton sphere. Here, $p_\parallel$ refers to the momentum along the polarization axis of the laser, and $p_\perp$ to be the momentum perpendicular to both the polarization axis and the spectrometer axis. We employ the polar onion-peeling algorithm \cite{roberts_toward_2009} to invert our raw spectrum, see the top half of FIG.~\ref{fig:Filtering}~(a).

In photoelectron spectra generated through SFI, the ATI rings tend to dominate and obscure other features present in the spectra. This is especially problematic for an analysis of holographic trajectory interferences in the direct ionization regime below 2U$_{\mathrm{p}}$ \cite{Becker2002}, where U$_{\mathrm{p}}$ is the ponderomotive energy of a free electron in the laser field \cite{bucksbaum_role_1987}. The ATI rings are a signature of multiple laser cycles, formed due to the interference of photoelectron pathways across these cycles. By removing these ATI rings from the spectrum, we can isolate the spectral features resulting from sub-cycle dynamics only. 

After inversion, we apply a time-filtering technique that effectively suppresses the contribution of inter-cycle interferences, particularly ATI rings, to the experimental PMD. The motivation and methodology for this technique is outlined in significantly more detail in a previous work \cite{werby_disentangling_2021}. In brief, the inversion process generates a set of one-dimensional anisotropy parameters dependent on the radial momentum $p_r$ which contain the full 3D information of the PMD \cite{reid_photoelectron_2003}. These parameters can be resampled to be functions of energy, which causes the ATI rings to be periodic. The reciprocal space of energy is time, so we are able to perform a low-pass Fourier filter on these resampled anisotropy parameters to suppress features caused by interfering electron trajectories which ionize at least one field cycle apart from each other. The result is shown in the top half of FIG.~\ref{fig:MainFig}~(b), where it is clear that ATI rings have been removed. A comparison between the inverted experimental data and the time-filtered data is shown in panel (a) of FIG.~\ref{fig:Filtering}.


We also apply this filtering procedure to the results of the CQSFA calculations. We first generate the photoelectron angular distribution (PAD) Legendre decompositions without onion-peeling to determine the anisotropy parameters for the CQSFA calculations.
Then we can filter the parameters using an identical filter to the one used for the experimental data to remove rapidly changing momentum features. Throughout the paper, everywhere we compare the CQSFA calculations directly to the experimental data, we filter them in this way. See FIG.~\ref{fig:Filtering}~(b) for a comparison of the CQSFA calculations with and without the filtering.
We note here that this paper serves as the first application of the time-filtering technique as a tool to make explicit measurements supporting quantum SFI theory.

\section{Unit-Cell Averaging}
\label{sec:unit_cell_averaging}


\begin{figure*}
	\centering
	\includegraphics[width=\linewidth]{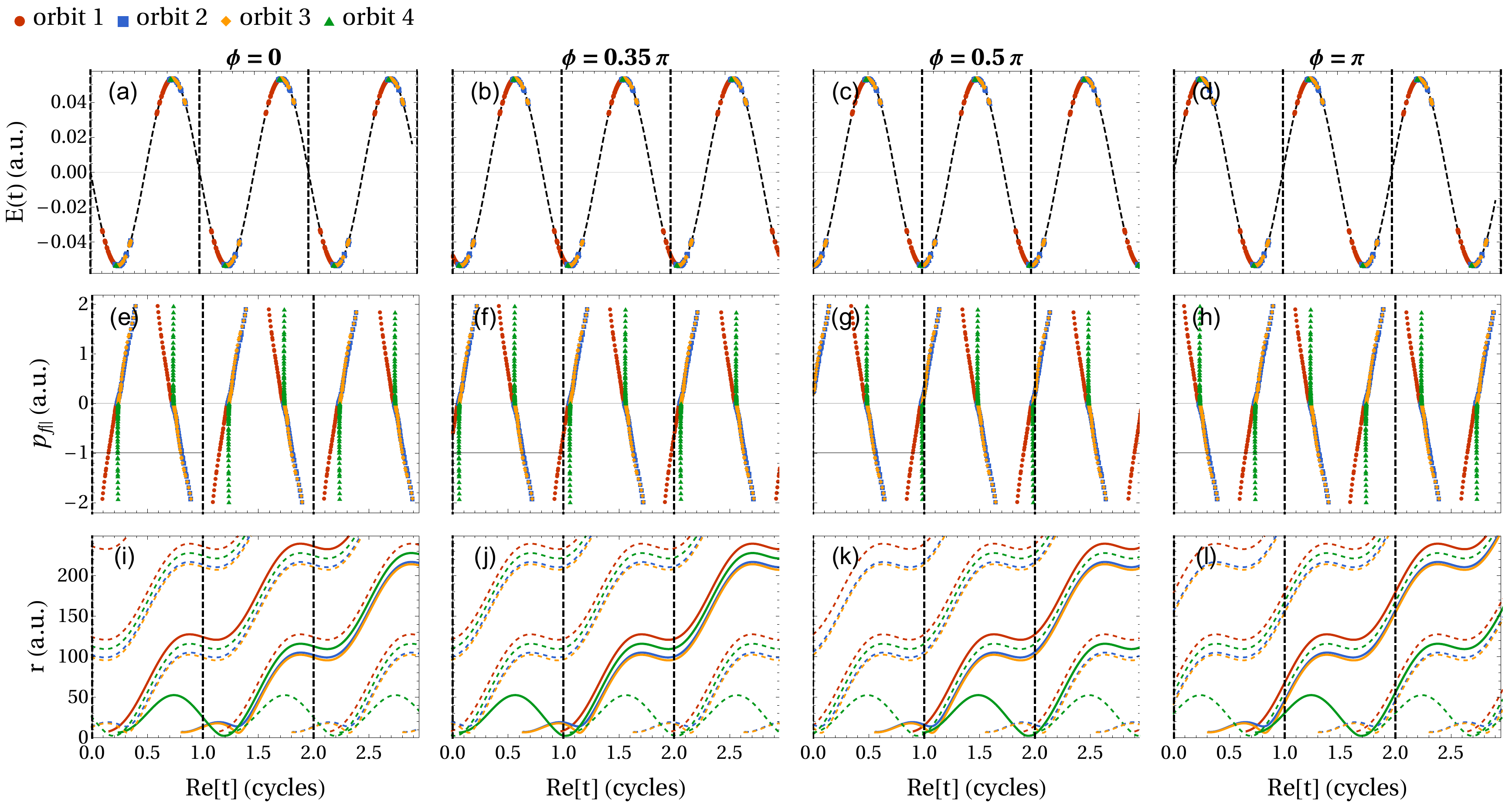}
	\caption{
	The periodic unit cell is exemplified in three ways: 1) In the top row by the monochromatic electric field over three cycles for four starting positions [(a)--(d)], denoted by $\phi$, with the time of ionization marked on the field for each CQSFA orbit. 2) The middle row displays the time of ionization vs the parallel final momentum, at a fixed perpendicular momentum of $p_{\perp}=0.13$~a.u.~, for all four orbits for the same four starting positions of the unit cell [(e)--(h)]. 3) The bottom row plots the distance from the parent ion over time for each CQSFA orbit for the same four starting positions [(i)--(l)]. The trajectories all have the final momentum of $\pb=(-1.0, 0.13)$~a.u. which is also marked by the horizontal line in the middle row. The solid trajectories indicate the ones that begin in/belong to the first unit cell, while the dashed trajectories belong to different unit cells. The unit cells are marked in all panels by vertical dashed lines. The line colors and markers correspond, in all panels, to the legend at the top.}
	\label{fig:CQSFATimes}
\end{figure*}

Here we discuss the key aspects of the CQSFA required to understand the unit-cell averaging methods employed. This method has been explored in detail in previous publications \cite{Lai2015, Lai2017,Maxwell2017,Maxwell2017a,Maxwell2018,Maxwell2018PRA,Maxwell2020,Kang2020,faria_it_2020} (see Refs.~\cite{Lai2015, Maxwell2017,faria_it_2020} for key details and a review), therefore, only a brief overview related to the present work is provided.

\begin{figure*}
    \centering
    \begin{subfigure}[b]{0.49\linewidth}
        \includegraphics[width=\linewidth]{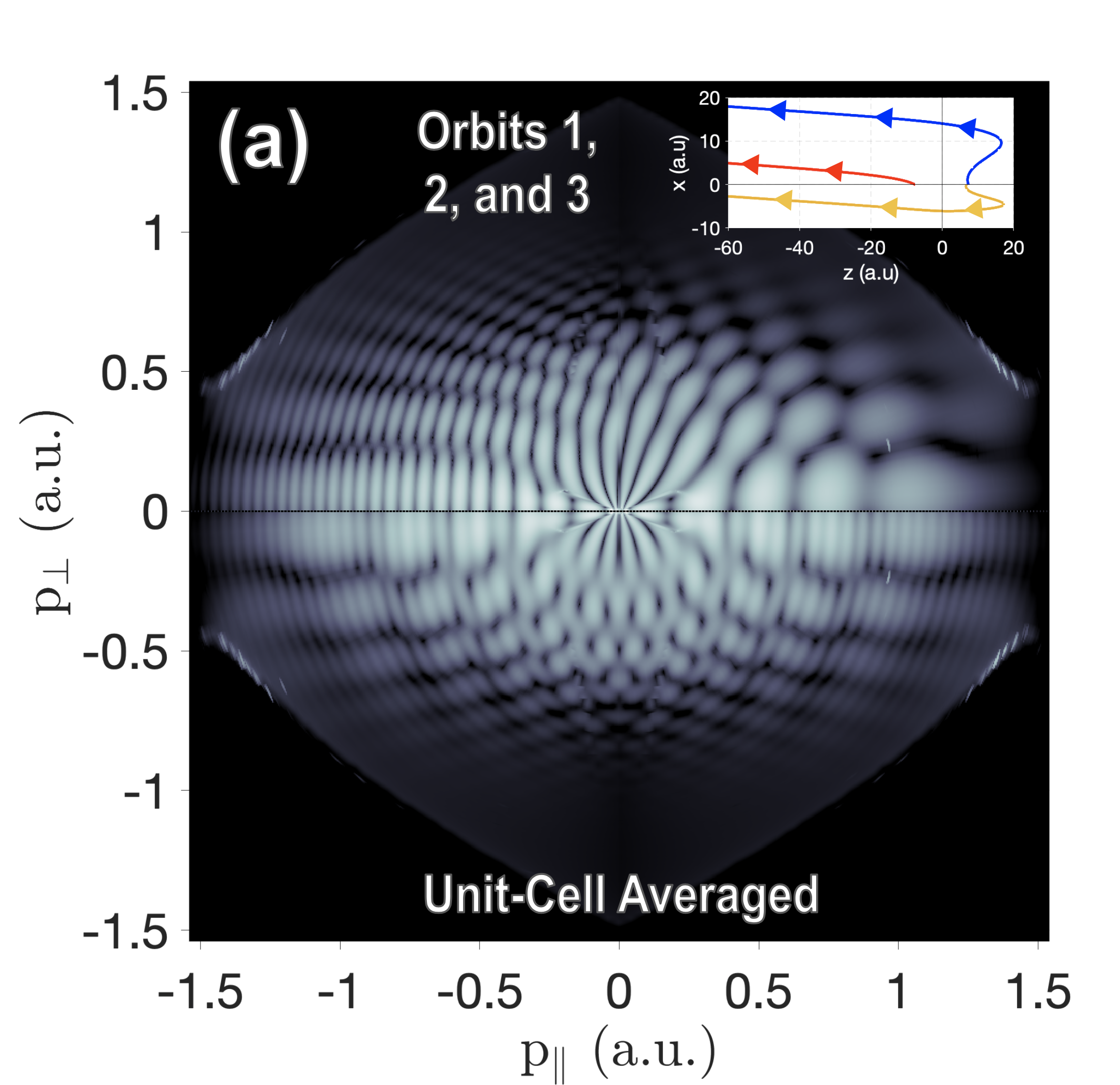}
    \end{subfigure}
    \begin{subfigure}[b]{0.49\linewidth}
        \includegraphics[width=\linewidth]{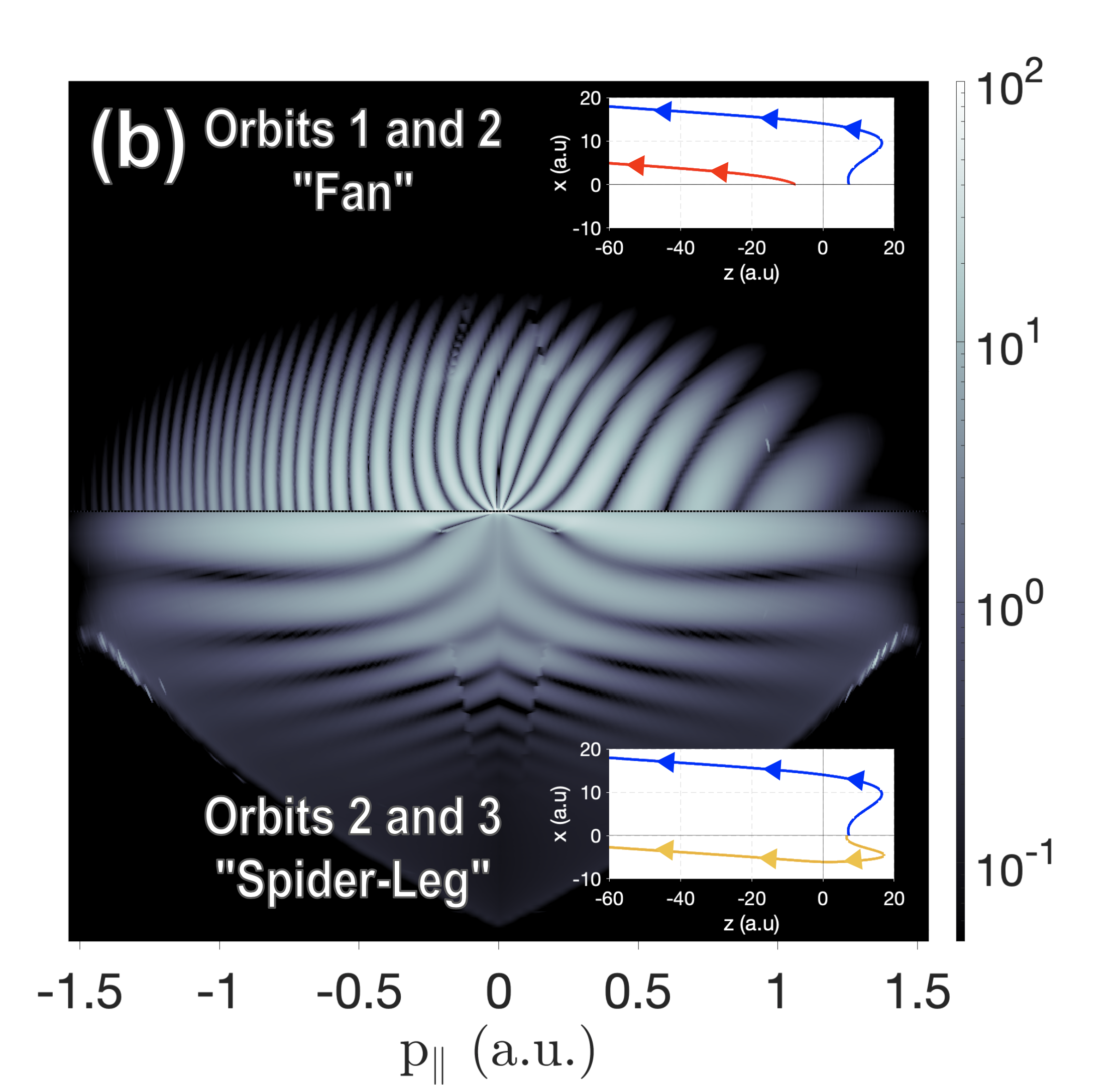}
    \end{subfigure}
    \caption{The origin of the modulations on the spider legs. The parameters are the same as in \figref{fig:MainFig}. Panel (a) shows the combined CQSFA calculation for three electron trajectories corresponding to orbits 1, 2, and 3 as presented in FIG.~\ref{fig:CQSFATrajecotories}. The bottom half of panel (a) shows the effect of unit-cell averaging as discussed in the text. Panel (b) displays the CQSFA computations including only orbits 1 \& 2 and 2 \& 3, in the top and bottom half, respectively.}
    \label{fig:ModulationOrigin}
\end{figure*}

In the CQSFA to model the electron dynamics within a single-cycle unit cell we employ a monochromatic field given by the vector potential
\begin{equation}
 \Ab(t) =2\sqrt{\up} \cos(\omega t+\phi), 
 \label{eq:AField}
\end{equation}
where $\omega$ is the angular frequency of the laser and the electric field is given by $\mathbf{E}(t)=-\partial \Ab(t)/ \partial t$. Note we employ atomic units throughout unless otherwise stated.
The variable $\phi$ is only important when the times are restricted to a single-cycle unit cell, where it controls the `starting position' of the laser field in the unit cell.
Importantly, all the electron dynamics are contained within the action. This is achieved by applying Feynman path integral formalism \cite{Kleinert2009} to the exact formalism of the transition amplitude given in Ref.~\cite{Becker2002}. With the application of the saddle point approximation this leads to the following expression for the ATI transition amplitude
\begin{align}
&M(\mathbf{p}_f)\propto\notag\\
&-i \lim_{t\rightarrow \infty } \sum_{s}\bigg\{\det \bigg[  \frac{\partial\mathbf{p}_s(t)}{\partial \mathbf{r}_s(t_s)} \bigg] \bigg\}^{-1/2} \hspace*{-0.6cm}
\mathcal{C}(t_s) e^{i
	S(\mathbf{p}_s,\textbf{r}_s,t,t_s))}
\label{eq:MpPathSaddle}
\end{align}
where 
\begin{equation}
\label{eq:Prefactor}
\mathcal{C}(t_s)=\sqrt{\frac{2 \pi i}{\partial^{2}	S(\mathbf{p}_s,\textbf{r}_s,t,t_s) / \partial t^{2}_{s}}}\langle \mathbf{p}+\mathbf{A}(t_s)|H_I(t_s)|\Psi_{0}\rangle,
\end{equation}
$s$ denotes the quantum orbits that solve the saddle point equations (see \eqref{eq:s1} and \ref{eq:s2}), which are summed over. There are four distinct types of orbits in the CQSFA, which will be described in more detail in \secref{sec:orbits}. The combination of these orbits leads to the interference patterns observed in \figref{fig:MainFig}. The interaction Hamiltonian is given by $\hat{H}_I(t)=-\hat{\mathbf{r}}\cdot \mathbf{E}(t)$. The action along each orbit reads 
\begin{equation}\label{eq:stilde}
    S(\mathbf{p},\textbf{r},t,t')=\ip t'-\int^{t}_{t'}[
    \dot{\mathbf{p}}(\tau)\cdot \mathbf{r}(\tau)
    +H(\mathbf{r}(\tau),\mathbf{p}(\tau),\tau)]d\tau,
\end{equation}
where $\ip$ is the ionization potential, the Hamiltonian 
$H(\rb(\tau),\pb(\tau),\tau)=1/2(\pb(\tau)+\Ab(\tau))^2+V(\rb(\tau))$ and $V(\rb)$ is given by the effective potential for argon previously employed in Refs.~\cite{Kang2020,Tong2005}.
An additional $-\pi/2$ shift is in specific cases added to \eqref{eq:stilde} to incorporate Maslov phase shifts not accounted for by using a 2-dimensional semi-classical model for a 3-dimensional system. This phase is added for every sign change in $p_{\perp}(\tau)$, as detailed in Ref.~\cite{Brennecke2020}, see \secref{sec:phases} for more details.
The momentum $\mathbf{p}$ and coordinate $\mathbf{r}$ have been parameterized in terms of the time $\tau$.
In this monochromatic field approximation, the actions are periodic in the variable $t'$. Thus, for any time of ionization $t'=t_s$, there are additional solutions $t'=t_s+ n T$, where $T$ is the period of the laser field and $n$ is any integer. Visualizations of the repeated trajectories are shown in \figref{fig:CQSFATimes}. 
The periodic ionization times across many cycles lead to the well-known ATI peaks/intercycle interference \cite{freeman_above-threshold_1987, Arbo2010, Arbo2012, Maxwell2017}, which using this approach, is described by an analytic formula and can be completely factored out \cite{Maxwell2017}.
The inter-cycle interference is not of interest for photoelectron holography as it does not add any extra information on the target. 
In fact, the ATI ring interference acts to obfuscate the holographic interference, so in these results we restrict the CQSFA ionization times to a single-cycle unit cell.
Restricting the ionization times but not the final propagation time allows physical processes that would be present in a real laser pulse and require multiple cycles, such as recollisions \cite{Becker2018}, to be approximated by the monochromatic theory, while removing the inter-cycle effects. This approach is an approximation to a real laser pulse, which neglects the laser envelope effects, but still can give very good agreement with experiment in the long-pulse case \cite{Kang2020, Maxwell2020}. Note this is not the same as using a single-cycle top-hat laser pulse, which would introduce radical switch on/off effects in the electron dynamics. A top hat pulse would also limit the possible processes, e.g. no electrons ionized in the second half cycle would return. Furthermore, it is not a realistic pulse to implement in the lab.

The periodic nature of the monochromatic CQSFA can be seen in \figref{fig:CQSFATimes}, where in panels (a)--(d) the laser field is plotted for different starting positions $\phi$ and the resulting times of ionization are marked on the field for each CQSFA orbit. The same ionization times are plotted directly below, panels (e)--(h), where the vertical axis displays the parallel final momentum to which each point corresponds. The perpendicular final momentum is fixed at $p_{\perp}=0.13$~a.u. The periodic nature is very clear over the 3 cycles plotted. As the `starting position', $\phi$, is increased the laser field and the times of ionization all shift to the left. This leads to earlier times of ionization leaving the first unit cell (marked by vertical dashed lines), while other times of ionization from the second unit cell move into the first. Thus, a different subset of trajectories are selected. This is shown explicitly in \figref{fig:CQSFATimes} (i)--(l), where the distance of each trajectory from the parent ion is plotted over time. The trajectories that have their starting time (i.e. ionization time) in the first unit cell are denoted with solid lines. These clearly change as the $\phi$ increases and different trajectories have their starting point in the first unit cell.

The variable $\phi$ has no bearing on the physics and different values will lead to the same symmetric momentum distribution if the full (infinite) duration of the monochromatic field is considered. Each unit cell represented in \figref{fig:CQSFATimes} contains the same information on the electron dynamics regardless of the value of $\phi$.  However, if considering only a single-cycle unit cell, the different ordering of the orbits [see \figref{fig:CEP_Avg} (a)-(d)] and discontinuous cuts through the ionization times of the orbits in momentum space [see \figref{fig:CEP_Avg} (e)-(h)] leads to asymmetry and discontinuities (where the unit cell `cuts' an orbit) in the final momentum distributions that change with $\phi$. 

As previously stated we wish to focus only on a single unit cell in order to examine the holographic sub-cycle effects, while disposing of the non-holographic intercycle interference. In the experiment we would like to simulate, the laser has a relatively long and gradually changing envelope. Furthermore, the CEP will vary from pulse to pulse, which will lead to different ordering of ionization pathways just like when $\phi$ is varied in the CQSFA. In the experiment we therefore expect that the measured photoelectron spectrum results from an incoherent average over all the allowed ordering (in time) of the ionization pathways.
Thus, an incoherent average of the momentum distribution with respect to $\phi$ in the CQSFA will combine the trajectories in different orders, as will be the case in the experiment, which will result in the removal of the asymmetries and discontinuities.

In the appendix we describe in detail how this can be achieved via integration over $\phi$. Here we present the unit-cell averaged probability $Prob(\pb)$, in terms of a `correction' to $Prob(\pb,0)=|M(\mathbf{p}_f)|^2$, the probability for $\phi=0$
\begin{align}
Prob(\pb_f)&=Prob(\pb_f, 0)
    +\frac{2\omega }{\pi}\sin\left(\Delta S/2\right)\times\notag\\
    &\hspace{1cm}\sum_{i<j} \Delta t_{ij} \Im\left[
    M_{i}(\pb_f)\overline{M_{j}(\pb_f)}e^{-i s_{ij}\Delta S/2}
    \right].
\end{align}	
Here, $M_i(\pb_{f})$ is the transition amplitude for $\phi=0$ for a CQSFA orbit $i\in[1,4]$, $\Delta S$ is a phase given by
\begin{equation}
    \Delta S = \frac{2\pi}{\omega}\left( 
            \ip + \up +\frac{1}{2}\pb_f^2
    \right),
\end{equation}
$\Delta t_{ij}=\Re[t'_i-t'_j]$ is the difference between the real part of the time of ionization of CQSFA trajectories $i$ and $j$ in the first unit cell for $\phi=0$ and $s_{ij}=\sign(\Delta t_{ij})$.

\section{Understanding Interference with Orbit-Based Model}
\label{sec:orbits}

Now that we understand how the experiment and theory can be brought together to disentangle sub-cycle interference, we exploit the ability of the CQSFA to turn on/off interference pathways to demonstrate the origin of the modulations on the spider-like interference patterns. 
To do this, we present some additional details of the CQSFA. Specifically, it is important to understand the four CQSFA trajectories, examples of which are given in \figref{fig:CQSFATrajecotories}. The equations of motion of the CQSFA trajectories are derived from the action via the application of the saddle point approximation, which leads to the saddle point equations
\begin{align} 
[\textbf{p}(t')+\textbf{A}(t')]^{2}/2+\ip=0,
\label{eq:s1}
\end{align}

\begin{align}
\dot{\pb}(\tau)=
-\nabla_\textbf{r}V[\rb(\tau)]
\ \text{ and } \
\dot{\rb}(\tau)= \pb(\tau)+\Ab(\tau).
\label{eq:s2} 
\end{align}
The first of these, \eqref{eq:s1}, provides the ionization times, while the pair of equations given by \eqref{eq:s2} describes the propagation in the continuum. The result is the four orbits shown in \figref{fig:CQSFATrajecotories}. These have been explained in detail in Refs.~\cite{Lai2015, Lai2017,Maxwell2017,Maxwell2017a,Maxwell2018,Maxwell2018PRA,Maxwell2020,Kang2020,faria_it_2020} (see Ref.~\cite{Maxwell2018} for the first implementation of all four orbits and Ref.~\cite{faria_it_2020} for a review) but a brief description follows.

The four orbits were originally classified in Ref.~\cite{Yan2010}, 
and they are shown in real space in \figref{fig:CQSFATrajecotories} for a specific final momentum.
Orbit 1 (direct): the electron tunnels towards the detector and reaches it directly. 
Orbit 2 (forward deflected) and orbit 3 (forward scattered): the electron tunnels away from the detector and then the laser drives them back towards the detector. For orbit 3 the electron's transverse momentum changes sign, for orbit 2 it does not. Orbit 4 (backscattered): the electron is freed towards the detector, but backscatters off the core.

Using combinations of these CQSFA orbits we can further investigate the interferences presented in \figref{fig:MainFig}. The modulations on the spider legs (interference between orbits 2 and 3) can be traced to the fan-like interference pattern (interference between orbits 1 and 2). 
In \figref{fig:ModulationOrigin} (a) we plot all three of these orbits (1, 2, and 3); in the bottom half of the panel we have applied unit-cell averaging and the modulations are clearly reproduced without requiring the inclusion of orbit 4. Unit-cell averaging has not been applied in the top half of the panel. We find this separates the modulations into fine modulations on the left and broad modulations on the right.
On the left and right sides of the top half of \figref{fig:ModulationOrigin} (b) we investigate the different modulations by plotting the fan, which is the interference between the two direct-like CQSFA orbits 1 and 2. The figure is asymmetric as no unit-cell averaging has been used ($\phi=0$). The two different interference types, seen in panel (a), are present on each side. In a previous publication \cite{Maxwell2017} we have referred to this as type A and B interference. Type A [B], relating to the broad [fine] fringes on the right [left], occurs when there is less [more] than half a cycle difference between the times of ionization of the two interfering electron pathways. 

\begin{figure}
    \centering
    \begin{subfigure}[b]{0.93\linewidth}
        \includegraphics[width=1\linewidth]{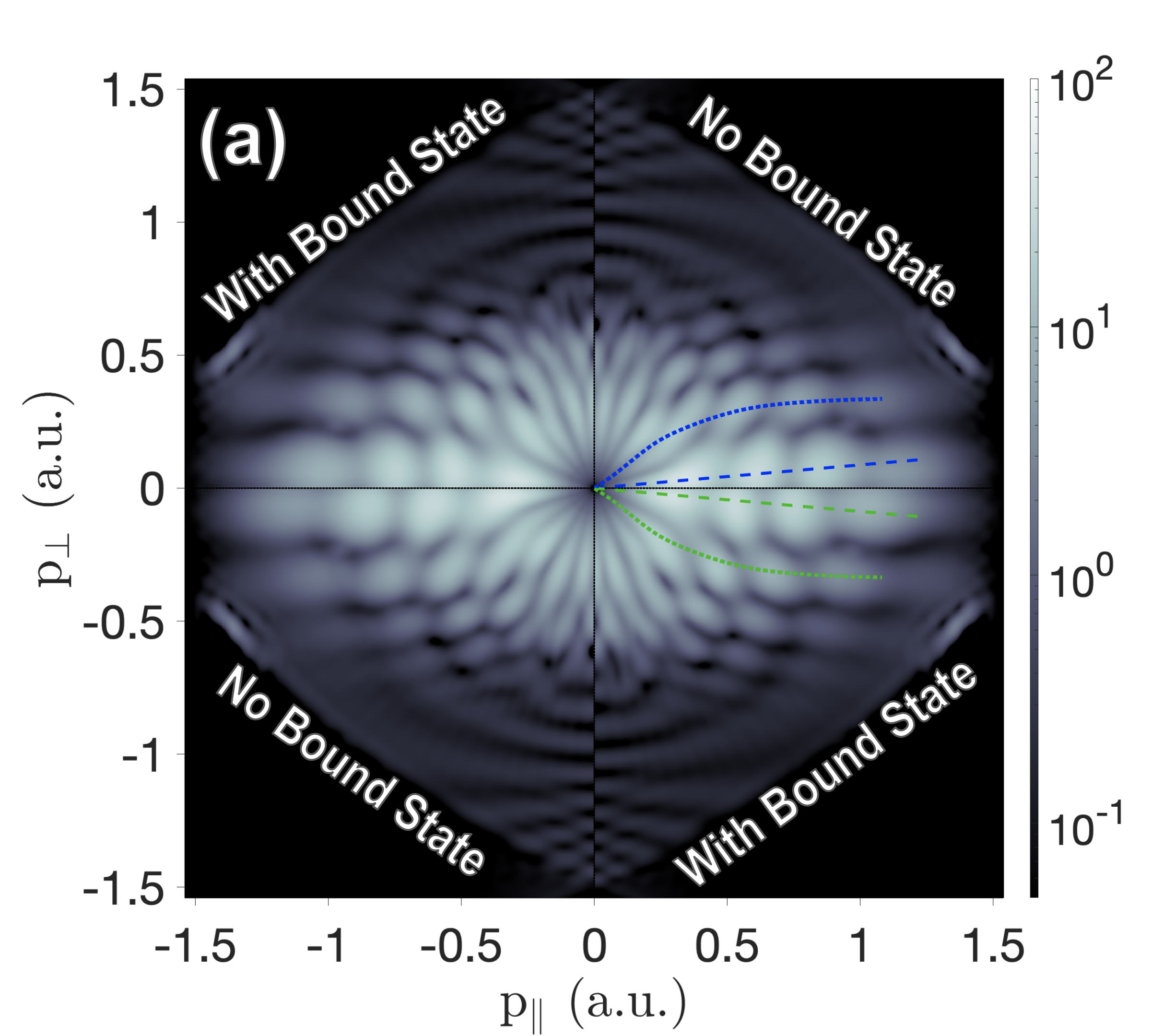}
    \end{subfigure}
    \vfil
    \begin{subfigure}[b]{0.93\linewidth}
        \includegraphics[width=1\linewidth]{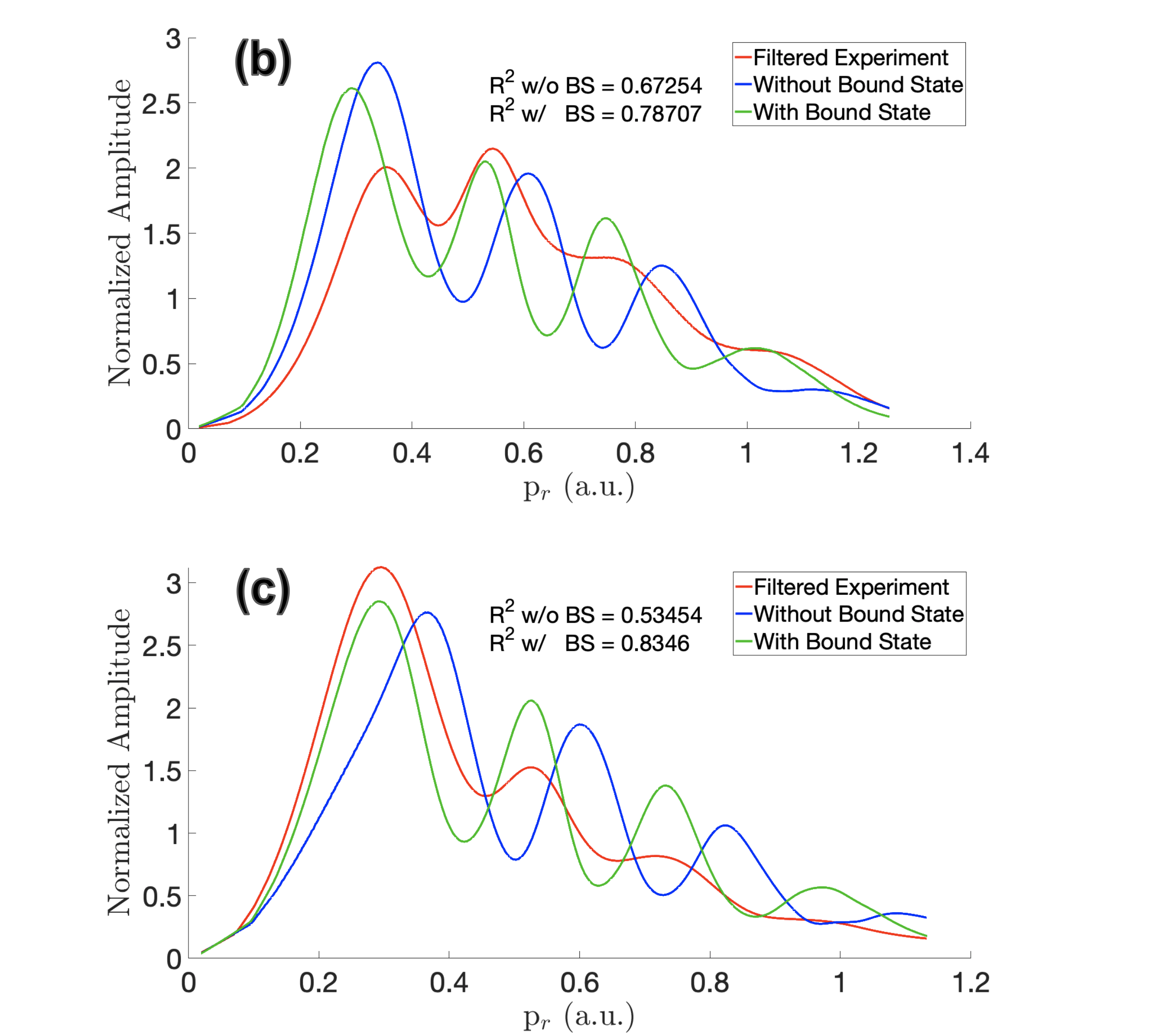}
    \end{subfigure}
    \caption{PMD computed using the CQSFA examining the effect of the bound state. The same parameters are used as in \figref{fig:MainFig}. Panel (a) displays the CQSFA with and without the effect of the bound state in alternating quadrants to enable the phase shift to be identified along both axes. The CQSFA PMDs have been filtered to remove high frequency structures. Panels (b) and (c) compare the lineouts along close to the parallel axis and along the first spider leg respectively as in FIG. \ref{fig:MainFig}. The same filtered experiment lineouts from FIG. \ref{fig:MainFig} are reproduced. The goodness of fit metric R$^2$ comparing the experiment lineouts to each of the other two in each plot is displayed. Note the bound state is `switched off' by setting the matrix element in \eqref{eq:Prefactor} to 1.} 
    \label{fig:ModulationPhase}
\end{figure}

We have fixed the laser field `starting position' $\phi=0$ such that, in \figref{fig:ModulationOrigin} (b), both type A and B occur on the right and left of the fan, respectively.
Unit-cell averaging will incoherently mix both interference types, however type A will tend to dominate.
If both interference types on the left and right of the top half \figref{fig:ModulationOrigin} panel (b) are added onto the spider legs in the bottom half of the panel then we get the results shown in \figref{fig:MainFig} and \figref{fig:ModulationOrigin} (a).
This shows clearly that the modulation effect is due to the interference of three electron trajectories (CQSFA orbits 1, 2, and 3) as well as an incoherent mix of different interference types A and B. Thus, in theory, we are able to see both sides of the fan imprinted in the spider. This has interesting consequences for photoelectron holography. For the spider it is known that the two interfering electron trajectories leave from the same side of the target but take different routes to the detector (with opposite transverse momentum components), while for the fan the two interfering trajectories leave from opposite sides and have opposite longitudinal momenta. Thus, the fan and the spider probe in opposite directions, so the three-trajectory combination has the capacity to probe in both directions simultaneously. This could allow for holographic imaging of the bound state in both these directions.

\section{Revealing Gouy and Parity Phases}
\label{sec:phases}

\begin{figure}
    \centering
    \begin{subfigure}[b]{0.93\linewidth}
        \includegraphics[width=1\linewidth]{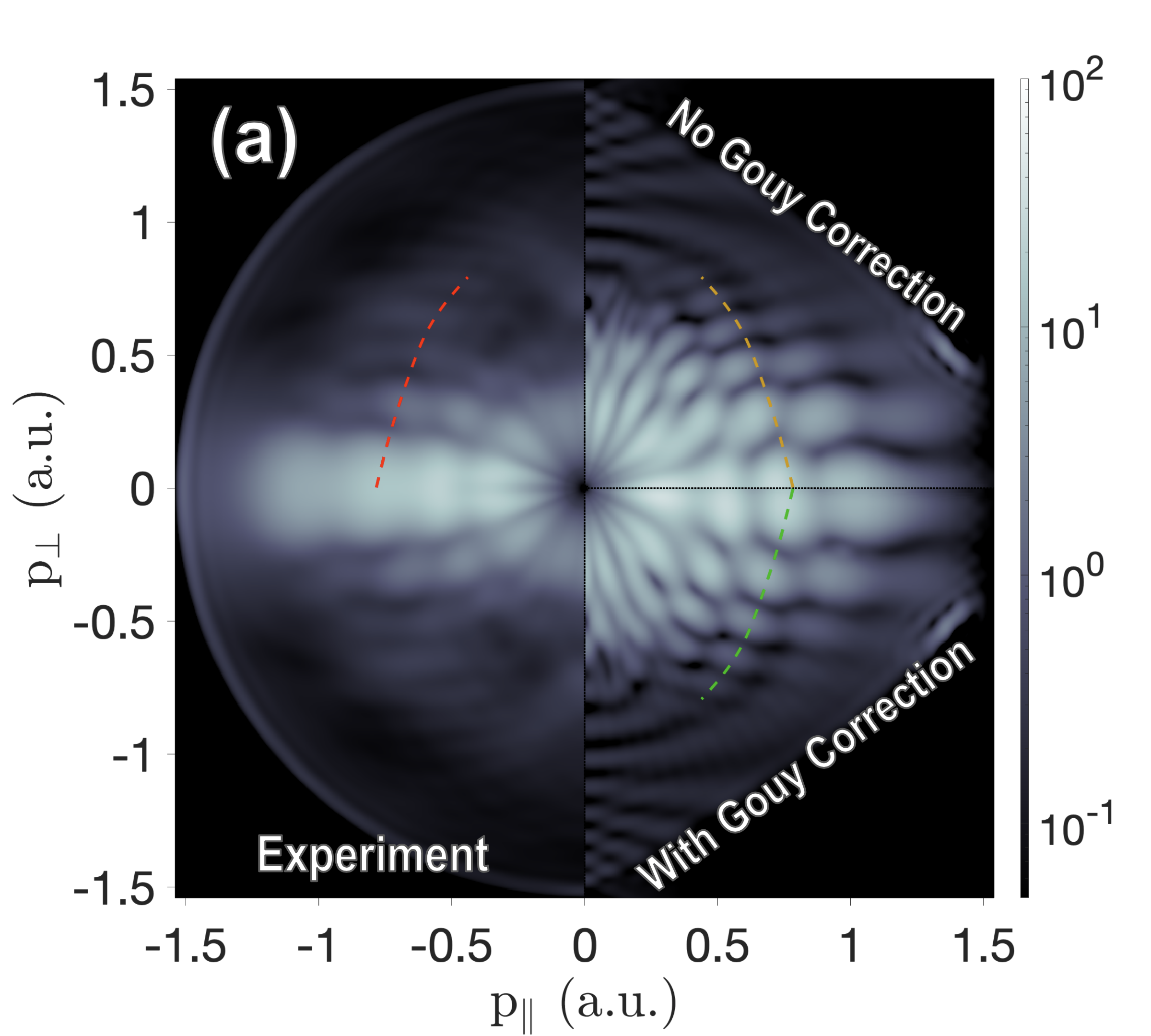}
    \end{subfigure}
    \vfil
        \begin{subfigure}[b]{0.93\linewidth}
        \includegraphics[width=1\linewidth]{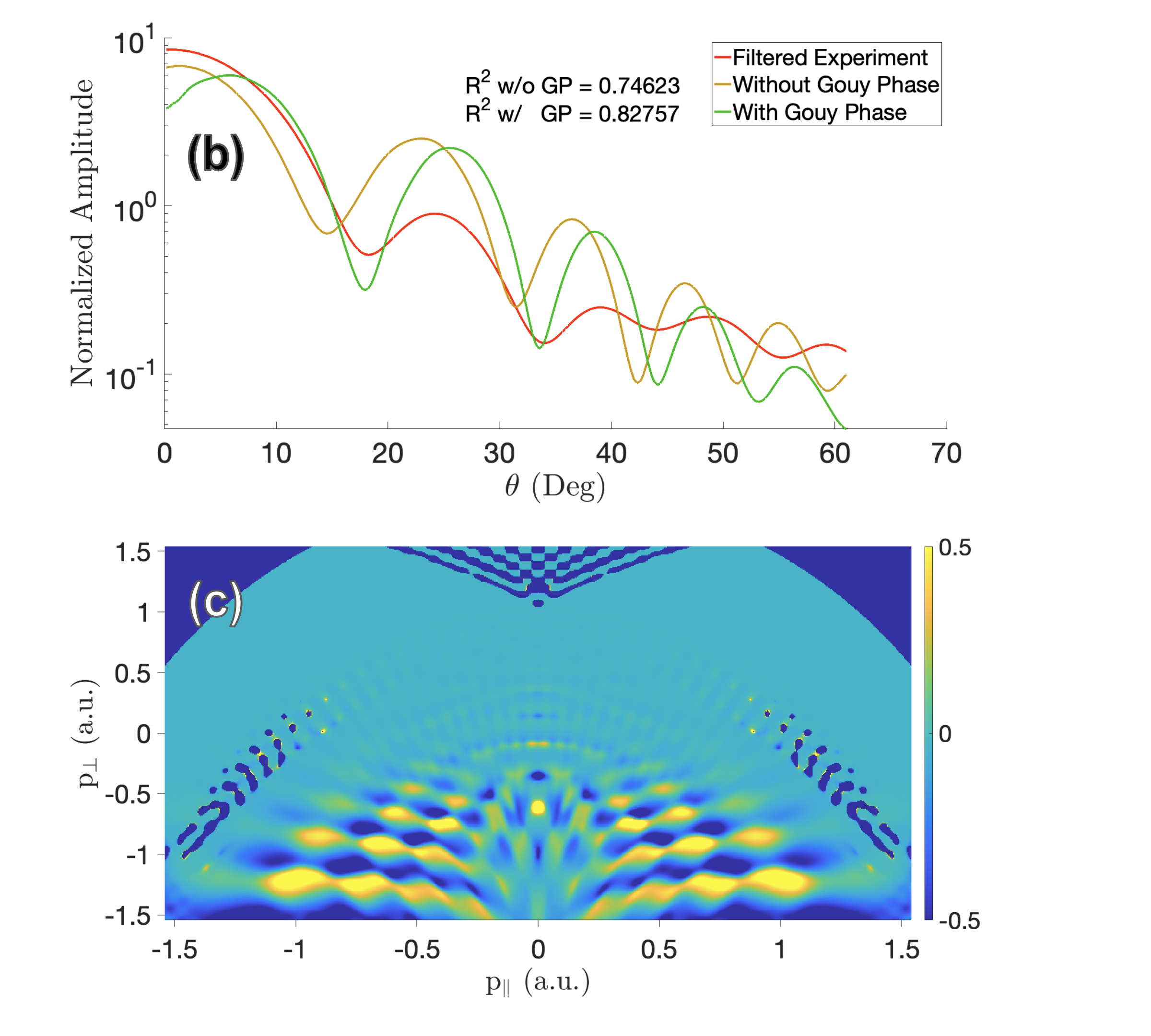}
    \end{subfigure}
    \caption{Comparison between the experimental data and the CQSFA computations with and without the Gouy phase correction (see text). Panel (a) displays the filtered experimental data and the CQSFA PMDs with and without the Gouy phase correction as labeled. 
    Panel (b) compares the lineouts taken from the measured maxima along the third antinode of the CQSFA computation with the Gouy phase correction as shown in panel (a).
    The goodness of fit metric R$^2$ is again shown, comparing the experiment to the CQSFA with and without the phase correction.  $\theta$ is measured equivalently for each lineout in panel (a) with the 0 at the axis and advancing along the lineout. Panel (c) displays the normalized residual of the CQSFA computation without the Gouy phase correction subtracted from the computation with the correction. This residual highlights the modification in the pitch of the spider-leg structure.} 
    \label{fig:GouyPhase}
\end{figure}

Previously, holographic interference has been used to probe parity in the bound state \cite{Kang2020}. This is possible as photoelectron trajectories that leave the ion from opposite sides will acquire an additional $\pi$ phase difference if the bound state orbital has odd parity, while there will be no additional phase difference for even parity. So for these trajectories the interference fringes will shift out of phase between odd and even parity.
In Ref.~\cite{Kang2020} only two trajectories were considered to extract the parity, primarily from the spiral-like structure, orbits 3 and 4. Furthermore a reference `atom' was required to use differential holographic measurements to extract the parity. The imprint of the fan in the spider-like structure allows us to see phase shifts between three trajectories. Here there will be a $\pi$ phase difference picked up between both orbits 1 and 2 as well as orbits 1 and 3.

We demonstrate the ability of probing the parity of bound state in \figref{fig:ModulationPhase} by adding and removing the effect of the odd parity $p$-state of argon. In panel (a) we plot with and without the effect of the bound state in alternating quadrants. Along the parallel axis it is clear that the fan-like modulations along the spider legs undergo a $\pi$ phase shift. This is due to the odd parity of the $p$-state of argon, so that trajectories leaving in opposite longitudinal directions pick up a $\pi$ phase difference. The same $\pi$ phase shift is also visible near the transverse axis at higher momenta via the spiral-like interference pattern, which occurs between forward- and back-scattered trajectories \cite{Maxwell2020}.

Lineouts traced along the parallel axis and the first spider leg are plotted in \figref{fig:ModulationPhase}~(b) and (c), along with the same experimental lineouts from \figref{fig:MainFig}. This comparison with experiment enables a direct corroboration of the $\pi$ phase difference due to the bound state. The $R^2$ `goodness of fit' is calculated in both cases of the CQSFA (with and without the bound state phases) vs the experiment. Along both the axial and first spider leg lineout including the bound state phases gives higher $R^2$ value. It is also evident that the peaks shift out of phase when the bound state phases are not included. Thus, we have demonstrated that this methodology can be used to determine phase inherent in the target.

In recent work \cite{Brennecke2020} it was demonstrated that additional Maslov phases (the semi-classical equivalent of Gouy phases) must be included to employ a 2-dimensional model for a 3-dimensional system, the additional phase for each trajectory is dependent on the number of sign changes of the perpendicular momentum $p_{\perp}(t)$.
In this case of the CQSFA these phases can be included by shifting the phase of orbits 3 and 4 by $-\pi/2$.

In \figref{fig:GouyPhase} we show the result of CQSFA computations with and without these phases as well as the experiment. It is particularly noticeable that the spider legs and axial fringes shift towards higher $p_{\perp}$ momentum. This leads to thicker fringes along the polarization axis and a steeper gradient along the spider legs, better matching experiment. The overall shift of the spider legs is exemplified in \figref{fig:GouyPhase}~(c), in which the normalized residual difference plot between the CQSFA with and without the additional Gouy-related phases is shown. In \figref{fig:GouyPhase}~(b) lineouts are shown for the CQSFA with and without the Gouy-related phases as well as the experiment. A much better match can be observed for the CQSFA with the Gouy phases, where the peaks almost line up with the experiment. In the case of the CQSFA without the Gouy phases, there is a constant phase shift away from the experiment. This provides further experimental verification of the additional phases predicted by Ref.~\cite{Brennecke2020}.

\section{Conclusions}
\label{sec:conclusions}

We present two new methods for bringing together experimental and theoretical results enabling an unprecedented quantitative match between the two. 
We have overcome two major obstacles to the interpretation of holographic strong-field ionization data: artificial defects present in theoretical models with restricted ionization times and strong inter-cycle ATI interference in experimental data.
This enables the identification of the first three-trajectory interference pattern in photoelectron holography, which has the capacity to strongly enhance current protocols. Such strong agreement also enables experimentally driven determination of intricate phases inherent within the system. Using a goodness-of-fit to the experiment we confirm that the bound state imparts a phase shift of $\pi$ on the CQSFA orbits 2 and 3, while the recently investigated Gouy phases \cite{Brennecke2020} (previously missing from the CQSFA computation) impart a phase shift of $-\pi/2$ on CQSFA orbits 3 and 4 due to potential focusing.

Previously, works on photoelectron holography have dealt with inter-cycle interference in different ways. In Refs.~\cite{he_direct_2018, meckel_signatures_2014}, short pulses were employed to suppress this interference. The issue with this is that it also acts to suppress some recoiling ionization pathways, which will encode the most information about the target. In Ref.~\cite{porat_attosecond_2018}, a longer pulse was employed and Fourier analysis was used to remove oscillations, but only for 1-D lineouts taken from the full 2D distribution. Additionally, no such analysis was applied to the corresponding theory employed therein.
In the seminal work on the spider-like structure \cite{huismans_time-resolved_2011} the ATI rings are visible in both the experiment and TDSE computation, 
which obscures the three-trajectory interference structure reported in this work.
Our methods provide a more complete analysis isolating sub-cycle holographic interference, without the drawbacks of previous works.

The time-filtering method developed to remove ATI rings from experimental PMDs \cite{werby_disentangling_2021} uses a Fourier transform analysis to remove inter-cycle interferences, which have a high frequency in energy space. In the experimental PMDs, this has the effect of suppressing interference patterns caused by interfering electron pairs ionized at least one field cycle apart. Likewise for the CQSFA calculations, unit-cell averaging restricts ionization to a single field cycle but considers the different time-ordering of trajectories resulting from alternative unit cells, allowing for the modelling of sub-cycle interference without asymmetry and discontinuities. 
It may seem redundant to use both methods upon the CQSFA calculations; however, the time-filtering technique removes more than just the ATI rings, with some finer features being subtracted as well. Ultimately, using both approaches clarifies the analysis, enabling all interference effects visible in experiment to be traced back using the CQSFA. The periodic nature of the CQSFA model is reminiscent of Floquet time crystals \cite{Else2016}, which motivates the idea of a unit cell and leads to unit-cell averaging. A recent review on Floquet analysis in materials in Ref.~\cite{Giovannini2019},
discusses period averaging that bears some resemblance to the unit-cell averaging. In Ref.~\cite{Roudnev2007} a similar technique is discussed for Floquet theory, which is referred to as CEP averaging.

The main benefit of unit-cell averaging is that it \emph{analytically} produces PMDs without ATI rings and takes into account all combinations of ionization pathways that occur in experiment. An alternative approach would mostly likely require two steps: firstly to model a host of laser pulses with different carrier envelope phases and then, secondly, to remove the ATI rings via the time-filtering technique in post processing. Not only would this take significantly more time to compute but it would be much harder to trace the origin of the final mixture of interference patterns. Our method is, of course, an approximation, which neglects the idea of a laser envelope; however, it yields precise agreement with experiment. As argued in the introduction, for the long pulses employed in this work, this will be a very good approximation to the electron dynamics. This argument of long pulses has been made before (e.g. Refs.~\cite{PhysRevLett.92.133006,PhysRevLett.119.243201,Maxwell2020,werby_disentangling_2021}) but in this work we significantly improve on this idea. 
Finally, the methods presented in this work are applicable to low and intermediate photoelectron energies, in which there is an intricate interplay of the binding potential, the external field, and the core dynamics. This, together with the high sensitivity of the methods, opens a wide range of possibilities for dynamical imaging of correlated multielectron systems in the attosecond regime. 


\begin{acknowledgments}
 NW, RF, and PHB thank James P. Cryan for useful and fruitful discussions. NW, RF, and PHB are supported by the U.S. Department of Energy, Office of Science, Basic Energy Sciences (BES), Chemical Sciences, Geosciences, and Biosciences Division, AMOS Program. ASM and CFMF are supported by funding from the UK Engineering and Physical Sciences Research Council (EPSRC). ASM acknowledges grant EP/P510270/1, which is within the remit of the InQuBATE Skills Hub for Quantum Systems Engineering.
CFMF would like to acknowledge EPSRC grant EP/T019530/1.

ASM also acknowledges support from ERC AdG NOQIA, Spanish Ministry of Economy and Competitiveness (``Severo Ochoa'' program for Centres of Excellence in R\&D (CEX\allowbreak{}2019-000910-S), Plan National FIDEUA PID2019-106901\allowbreak{}GB-I00/10.13039/501100011\allowbreak{}033, FPI), Fundació Privada Cellex, Fundació Mir-Puig, and from Generalitat de Catalunya (AGAUR Grant No.\ 2017 SGR 1341, CERCA program, QuantumCAT \_U16-011424, co-funded by the ERDF Operational Program of Catalonia 2014-2020), MINECO-EU QUANTERA MAQS (funded by State Research Agency (AEI) PCI2019-111828-2/10.\allowbreak{}13039/\allowbreak{}501100011033), EU Horizon 2020 FET-OPEN OPTOLogic (Grant No 899794), and the National Science Centre, Poland-Symfonia Grant No.\ 2016/20/W/ST4/00314.
\end{acknowledgments}

\appendix
\section{Unit-Cell Averaging in the CQSFA}
\label{appendix:unit_cell_averaging}
\begin{figure*}
    \centering
    \includegraphics[width=\textwidth]{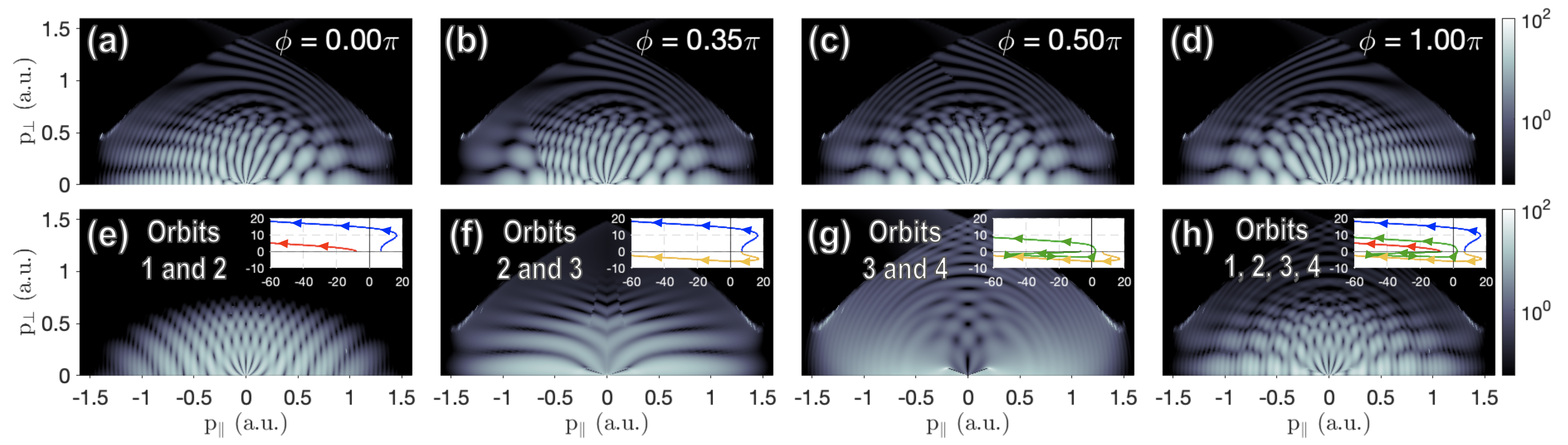}
    \caption{Top row shows PMDs computed using different values for $\phi$, which corresponds to those used in \figref{fig:CQSFATimes}. The bottom row shows unit-cell averaged PMDs with different combinations of CQSFA orbits indicated in each panel. The target and field parameters are the same as those used in \figref{fig:CQSFATimes}.} 
    \label{fig:CEP_Avg}
\end{figure*}

In \secref{sec:unit_cell_averaging} we outlined the ideas and main equations behind the new unit-cell averaging. In this section of the appendix we will fully derive the equation. We start by considering what happen when the starting phase $\phi$, introduced in \secref{sec:unit_cell_averaging}, is increased from $0$. As previously demonstrated some trajectories will move outside the unit cell. The time of ionization for an arbitrary $\phi$ can be written as $t'+\phi$, where $t'$ is the time of ionization for $\phi=0$. Thus, a trajectory will move out of the unit cell if the real part of the time is less than zero, which leads to the condition $\omega\Re[t']<\phi$. When this condition is satisfied the trajectory must be delayed by a field cycle in order for it to occur in the first unit cell. The delay amounts to including an additional phase $\Delta S$ given by
\begin{equation}
    \Delta S = \frac{2\pi}{\omega}\left( 
            \ip + \up +\frac{1}{2}\pb_f^2
    \right).
\end{equation}
With this in mind, we can now write an expression for the transition amplitude that is valid for any $\phi$
\begin{equation}
    M_{i}(\pb_f, \phi)=M_{i}(\pb_f)\exp\left[i H(\phi-\omega t^{Re}_{i}) \Delta S \right],
\end{equation}
where $M_{i}(\pb_f)$ is the transition amplitude, $i\in [1, 4]$ denotes the CQSFA orbit, $t^{Re}_{i}$ is the real part of the time of ionization for the CQSFA orbit at $\phi=0$ and $H$ is the Heaviside step function. 
The $\phi$ dependent probability distribution can be computed via
\begin{equation}
    Prob(\pb_f,\phi)=\left| 
            \sum_{i=1}^{4}M_{i}(\pb_f, \phi)
    \right|^2.
    \label{eq:prob_phi}
\end{equation}

With the definition given by \eqref{eq:prob_phi} we can plot the PMDs for the CQSFA at different values of $\phi$, this has been done in the top row of \figref{fig:CEP_Avg}.
The values $\phi=0$ and $\pi$ do not exhibit discontinuities in the PMDs. All values of $\phi$ in between these values will have a discontinuity, which will occur when a trajectory moves outside of the unit cell. 
The values $\phi = 0$ and $\pi$ result in asymmetric momentum distributions, which contradict the symmetry of the experiment, and are related by flipping the $p_{||}$ axis. They
exhibit two types of broad and fine interference, previously dubbed type A and B, respectively \cite{Maxwell2017}. The value $\phi=0.5\pi$ is nearly symmetric, with a curved discontinuity near $p_{||}=0$,  but it contains almost exclusively type A interference. For the case of $\phi=0.35\pi$ a diagonal discontinuity can be seen on the left of the panel.

In order to perform unit-cell averaging the probability distribution given by \eqref{eq:prob_phi} must be integrated over all possible values of $\phi$
\begin{align}
    Prob(\pb_f)&:=\frac{1}{2\pi}\int_0^{2\pi} d \phi Prob(\pb_f,\phi) \notag\\
    &=\frac{1}{2\pi}\int_0^{2\pi} d \phi\left|
            \sum_{i=1}^{4}M_{i}(\pb_f)\exp\left[i H(\phi-\omega t^{Re}_{i}) \Delta S\right]
     \right|^2,
     \intertext{which may be written as}
     Prob(\pb_f)&=\frac{1}{2\pi}\sum^4_{i,j=1} M_{i}(\pb_f)\overline{M_{j}(\pb_f)}
        I_{\phi}.
\end{align}
Here $I_{\phi}$ is given by
%
\begin{align}
    I_{\phi}&=\int_0^{2\pi} d \phi 
        \exp\left[i (H(\phi-\omega t^{Re}_{i})-H(\phi-\omega t^{Re}_{j})) \Delta S\right]\notag\\
        &=2\pi+\omega |\Delta t_{ij}|\left(e^{-is_{ij} \Delta S}  -1\right),
\end{align}
where $\Delta t_{ij}=t^{Re}_{i} - t^{Re}_{j}$ and $s_{ij}=\sign(\Delta t_{ij})$. Inserting this into the probability distribution yields
\begin{align}
    Prob(\pb_f)&=Prob(\pb_f, 0)\notag\\
    &+\frac{\omega}{2\pi}\sum_{i,j=1}^{4} M_{i}(\pb_f)\overline{M_{j}(\pb_f)}
     |\Delta t_{ij}|\left(e^{-i s_{ij} \Delta S}  -1\right).
     \intertext{With some algebra this becomes}
     Prob(\pb_f)&=Prob(\pb_f, 0)
    +\frac{2\omega }{\pi}\sin\left(\Delta S/2\right)\times\notag\\
    &\hspace{1cm}\sum_{i<j} \Delta t_{ij} \Im[M_{i}(\pb_f)\overline{M_{j}(\pb_f)}e^{i s_{ij}\Delta S/2}].
    \label{eq:unit-cell-averaging}
\end{align}
Thus, the unit-cell averaging can be seen as a `correction' to the probability distribution for $\phi=0$, which uses only the transition amplitude for $\phi=0$, the real parts of the time of ionization for the orbits and the additional phase $\Delta S$.

The bottom half of \figref{fig:CEP_Avg} shows PMDs where \eqref{eq:unit-cell-averaging} has been applied to perform unit-cell averaging for different combinations of the CQSFA orbits. \figref{fig:CEP_Avg} (e) shows unit-cell averaged orbits 1 and 2 (fan-like structure), which form the modulations on the spider via the incoherent combination of broad and fine interference from both sides of original PMD. In \figref{fig:CEP_Avg} (f) we show the spider-like interference (orbits 2 and 3), unit-cell averaging has no effect as the two trajectories have very similar ionization times hence $\Delta t_{ij}\approx 0$. We show the spiral-like structure in \figref{fig:CEP_Avg} (g); unit-cell averaging leads to the carpet-like structure \cite{Korneev2012,Korneev2012a,Maxwell2020} without requiring the addition of ATI rings. Finally, in \figref{fig:CEP_Avg} (h) all orbits with unit-cell averaging are shown as in \figref{fig:MainFig}.

\bibliography{main}{}
\bibliographystyle{apsrev4-2}

\end{document}